\documentclass[aps,prb,
10pt,
showpacs,
]{revtex4-1}
\usepackage{bm}
\usepackage{amsmath}    
\usepackage{amssymb}    
\usepackage{graphicx}   
\usepackage{booktabs}
\DeclareMathOperator{\sech}{sech}

\usepackage[hang]{subfigure}
\newcommand{\jun}{junction }
\newcommand{\juns}{junctions }
\newcommand{\Jos}{Josephson }

\begin{document}
\title[Roberto Monaco]{Josephson Vortex Qubit based on a Confocal Annular Josephson Junction}
\author{Roberto Monaco}
\email[Corresponding author e-mail address:]{  r.monaco@isasi.cnr.it and roberto.monaco@cnr.it}
\affiliation{CNR-ISASI, Institute of Applied Sciences and Intelligent Systems ''E. Caianello'', Comprensorio Olivetti, 80078 Pozzuoli, Italy}

\author{Jesper Mygind}
\affiliation{DTU Physics, B309, Technical University of Denmark, DK-2800 Lyngby, Denmark}
\email{myg@fysik.dtu.dk}

\author{Valery P.\ Koshelets}
\affiliation{Kotel'nikov Institute of Radio Engineering and Electronics,
Russian Academy of Science, Mokhovaya 11, Bldg 7, 125009 Moscow, Russia.}
\email{valery@hitech.cplire.ru}

\date{\today}

\begin{abstract}
We report theoretical and experimental work on the development of a Josephson vortex qubit based on a confocal annular Josephson tunnel junction (CAJTJ). The key ingredient of this geometrical configuration is a periodically variable width that generates a spatial vortex potential with bistable states. This intrinsic vortex potential can be tuned by an externally applied magnetic field and tilted by a bias current. The two-state system is accurately modeled by a one-dimensional sine-Gordon like equation by means of which one can numerically calculate both the magnetic field needed to set the vortex in a given state as well as the vortex depinning currents. Experimental data taken at $4.2\,K$ on high-quality $Nb/Al$-$AlOx/Nb$ CAJTJs with an individual trapped fluxon advocate the presence of a robust and finely tunable double-well potential for which reliable manipulation of the vortex state has been classically demonstrated. The vortex is prepared in a given potential by means of an externally applied magnetic field, while the state readout is accomplished by measuring the vortex-depinning current in a small magnetic field. Our proof of principle experiment convincingly demonstrates that the proposed vortex qubit based on CAJTJs is robust and workable.
\end{abstract}
\maketitle
\tableofcontents
\newpage
\listoffigures
\newpage

\section{Introduction}
\noindent According to quantum mechanics, a massive particle subjected to potential confinement has its energy quantized and a discrete energy spectrum would be expected in the classical region of positive kinetic energy. The energy levels can be probed by irradiating the system with microwaves that resonantly excite the particle from the ground state to the first excited states (microwave spectroscopy). Quantum tunneling allows the possibility to escape from a potential well, passing the classically forbidden region. Due to tunneling, the ground state in a double-well potential is a doublet with energy splitting which depends critically on the precise shape and scale of the potential and, when a measurement is made, the particle is found in one of the two possible states $|L\rangle$ or $|R\rangle$ with a probability that oscillates in time \cite{landau}. In an asymmetric double-well potential the linear superposition of the two nearly-degenerate macroscopically distinct ground states has become very important in quantum information theory. In fact, the coherent oscillation between the basis states is the key ingredient for the realization of an elementary bit of quantum information (qubits, i.e., two-state quantum-mechanical systems) capable of implementing quantum computing. The quantum superposition of the two basis states can be manipulated by resonant microwave pulses.

Several two-state superconducting devices based on different degrees of freedom have been experimentally demonstrated as viable solid-state qubits in analogy with atomic and molecular systems. Indeed, Rabi oscillations, namely the oscillations in the population of the first excited level as a function of the applied microwave power, which are a preliminary requirement of quantum computing, have been reported in so-called charge \cite{Makhlin,You}, flux \cite{Mooij99,Friedman00}, and phase \cite{Yu} qubits. The operation of these systems, once sufficiently decoupled from their environment, is based on quantum coherence of the charge state, the magnetic-flux state, or the Josephson phase state, respectively, in circuits made of low-capacitance \Jos Tunnel Junctions (JTJs). In distinction to atoms, superconducting qubits which are driven by static electric and magnetic fields, as well as microwave photons, are strongly coupled to the environment. They can be fabricated by established lithographic methods, and the preparation, manipulation and measurement techniques are relatively simple. In addition, their performance has improved by several orders of magnitude in the past decade. The continuing evolution of designs and operational principles demonstrates the robustness and future potential of the field.

A fourth type of superconducting qubit was implemented which exploits the coherent superposition of two spatially separated states for a Josephson vortex (a supercurrent loop carrying one magnetic flux quantum also called fluxon) within a long and narrow (planar) JTJ, the Josephson Transmission Line (JTL), in which the spatial degree of freedom gives rise to the existence of topological singularities (fluxons). JTLs are well suited systems for the experimental study of nonlinear waves existing in the sine-Gordon system. In the Josephson Vortex Qubit (JVQ), the center of mass of the fluxon becomes the macroscopic collective coordinate of a quantum ‘particle’ existing within a potential well which can contain discrete energy levels. As for all the other solid-state qubits, there exists a crossover temperature which separates the thermally activated region and the quantum tunneling region. At high temperature, in the classical regime, the fluxon can escape from a potential well, lifted by thermal energy over the barrier. At low enough temperature $T$ ($k_B T$ smaller than the energy splitting of the qubit) when most of the dissipative mechanisms are eliminated \cite{JAP92}, the quantum regime establishes and the fluxon escapes occurs by macroscopic quantum tunneling through the barrier \cite{devoret84}. This process can be resonantly activated by a weak microwave perturbation. Macroscopic quantum tunneling is important to test the validity of the quantum mechanics on scales larger than the atomic one \cite{leggett02}.

So far the fluxon quantum effects have been observed only in curved JTLs whose extremities are jointed to form a doubly-connected or annular JTL where the boundary conditions of the open simply-connected configuration are replaced by periodic conditions. A unique property of not simply-connected \juns is the fluxoid quantization \cite{mercereau63} in the superconducting loop formed by either the top or the bottom electrode of the tunnel junction. Then, one or more fluxons may be topologically trapped in the \jun during the normal-superconducting transition. The existence of quantized levels of the vortex energy within the trapping potential well was demonstrated by measuring the statistics of the vortex escape from a magnetically-induced pinning potential in a $0.5\, \mu m$-wide ring-shaped JTL at temperatures below $100\,mK$ \cite{wallraff00}; later on, the vortex quantum tunneling was reported in a spatially dependent potential tailored by locally changing the radius of curvature of the annular junction \cite{goldobin01} to form a heart-shaped JTL \cite{wallraff03}. However, the coherent oscillation between the basis states, the key ingredient for the realization of a qubit, has not yet been observed for JVQs. In both cases the potentials were induced by an externally applied uniform magnetic field. The two macroscopically distinct quantum states needed for the JVQ may also be created by local magnetic fields induced by control currents \cite{ustinov02,fistulPRB03} or even by residual spurious fields \cite{carapella04}. Other vortex qubit prototypes were suggested in which the double-well potential is produced by two closely implanted defect sites in the insulator layer \cite{kato96,kim06} or by two artificially created discontinuities of the Josephson phase \cite{Heim13}.

\begin{figure}[b]
\centering
\includegraphics[height=6cm]{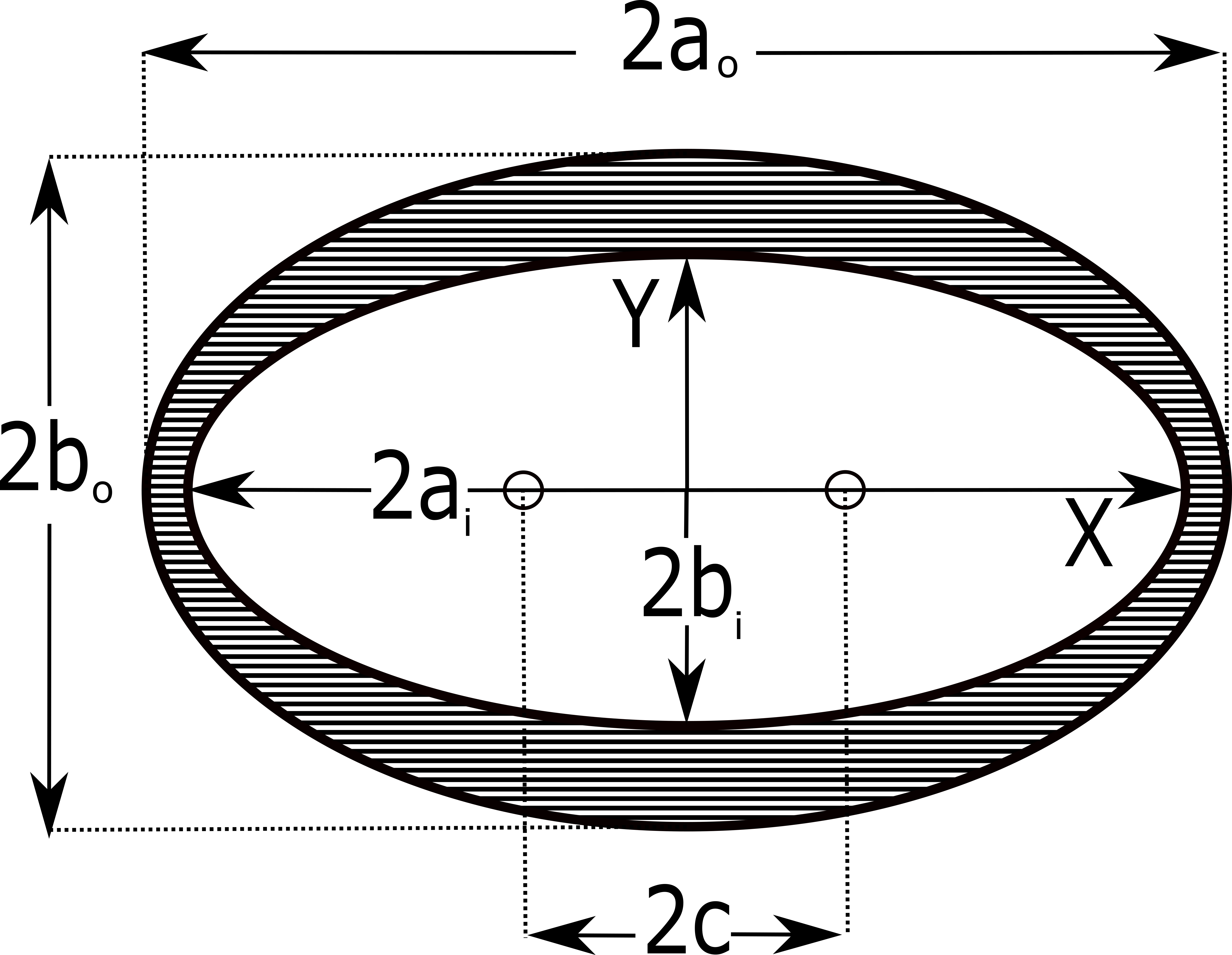}
\caption{Drawing of a confocal annulus delimited by two closely spaced confocal ellipses, representing the tunneling area of a CAJTJ. The two open circles are the common ellipses foci. The annulus width is smallest at the equatorial points and largest at the poles.}
\label{ConfAnn}
\end{figure}

In studying the Josephson vortex ratchet potentials, Goldobin \textit{et al.} found that for a variable-width JTL, as far as the width does not change much over the distance compared to the fluxon size, the fluxon potential just repeats the width profile \cite{goldobin01}. It follows that a large variety of spatially dependent fluxon potentials can be engineered in JTLs having a non-uniform width, provided that the width-dependence of the fluxon rest mass \cite{kato96,ustinov06} is taken into account in the kinetic energy. Indeed, the existence of a fluxon repelling (attracting) barrier induced by a slowly widening (narrowing) JTL \cite{benabdallah96} has been recently investigated \cite{JPCM16} to form a magnetically tunable double-well potential in variable-width annular JTLs named Confocal Annular Josephson Tunnel Junctions (CAJTJs)\cite{JLTP16b} since their tunneling area is delimited by two closely spaced ellipses having the same focal length; the tunneling area of a CAJTJ is shown in Figure~\ref{ConfAnn} where the principal diameters of the ellipses, $2a_i$ and $2b_i$ for the inner one and $2a_o$ and $2b_o$ for the outer one, are made parallel to the $X$ and $Y$ axes of a Cartesian coordinate system and the common foci $(\pm c,0)$ lie on the $X$-axis. The width of the confocal annulus is smallest at the equatorial point, $\Delta w_{min}=a_o-a_i$, and largest at the poles, $\Delta w_{max}=b_o-b_i$; the width variation is smoothly distributed along one fourth of the JTL perimeter. It is this smooth periodic change of the annulus width that makes the physics of CAJTJs very rich and interesting and the modeling very accurate. As the ellipses foci move towards the origin, the annulus eccentricity vanishes and the confocal annulus progressively reduces to a circular annulus with uniform width. Such ring-shaped JTLs were recognized to be ideal devices not only to experimentally test the perturbation models developed to take into account the dissipative effects in the propagation with no collisions of sine-Gordon kinks \cite{davidson85,dueholm,hue}, but also to investigate both the static and the dynamic properties of fluxons in the spatially periodic potential induced by an in-plane magnetic field,\cite{gronbech, ustinov,PRB98,wallraff03}. The potential felt by a fluxon trapped in a CAJTJ follows the variation of the width with minima (maxima) at the equatorial (polar) points. In addition, a large variety of fluxon potentials can be constructed by tuning the externally applied magnetic field and bias current. The aim of this work is to provide the experimental evidence of this potential in the thermal (or classic) regime, with emphasis on the preparation and readout of the vortex state. We show that the manipulation of the vortex state can be accomplished by means of either a barrier-parallel or transverse magnetic field.



\subsection{ Outline of the paper}

\noindent The paper is organized as follows. In Sec. II, we state the problem by describing the geometrical properties of a CAJTJ and introduce the mathematical notations and identities used throughout this paper. In addition, we review the modeling framework of our study, which is based on a modified and perturbed sine-Gordon equation, and provide the expression of the kinetic and potential energies for a fluxon trapped in a current-biased CAJTJ subjected to an external magnetic field. In Sec. III we present numerical simulations concerning the fluxon static and dynamic properties in underdamped CAJTJs and describe a protocol to reliably prepare and determine the vortex state. In Sec. IV we describe the experimental setup, the fabrication of our high-quality low-loss $Nb/Al$-$AlOx/Nb$ window Josephson tunnel junctions and the geometries that have been realized; later on, we present the experimental data of CAJTJs with both in-plane and transverse magnetic fields and discuss the role of the magnetic self-effect as well as of a non-uniform current distribution. Finally, a characterization of the two-state potential in the thermal (or classic) regime is presented in Section V with emphasis on the preparation and readout of the vortex state. The conclusions are drawn in Section VI.

\section{ Theory of one-dimensional CAJTJs} 

The geometry of our system suggests the use of the (planar) elliptic coordinate system $(\nu,\tau)$, a two-dimensional orthogonal coordinate system in which the coordinate lines are confocal ellipses and hyperbolae. In this system, for a given positive $c$ value, any point $(x,y)$ in the $X$-$Y$ plane is uniquely expressed as $(c\cosh\nu\sin\tau, c\sinh\nu\cos\tau)$ with $\nu\geq0$ and $\tau\in[-\pi,\pi]$. According to these notations, the origin of $\tau$ lies on the positive $Y$-axis and increases for a clockwise rotation. In the limit $c\to0$, the elliptic coordinates $(\nu,\tau)$ reduce to polar coordinates $(r,\theta)$, where $\theta$ is the angle relative to the $Y$-axis; the correspondence is given by $\tau\to \theta$ and $c\cosh\nu\to r$ (note that $\nu$ itself becomes infinite as $c\to0$). Once the foci position is given, all the possible confocal ellipses are uniquely identified by a value of $\nu$; we will name $\nu_i$ and $\nu_o>\nu_i$ the characteristic values of, respectively, the inner and outer CAJTJ boundaries. Their mean value, $\bar{\nu}= (\nu_o+\nu_i) /2$, labels one more confocal ellipse in between, called mean or master ellipse with principal axes $\bar{a}=c \cosh\,\bar{\nu}$ and $\bar{b}=c \sinh\,\bar{\nu}$ such that $a_i<\bar{a}<a_o$ and $b_i<\bar{b}<b_o$. A confocal annulus is said to be narrow when $\Delta\nu\equiv \nu_o-\nu_i=(a_o-a_i)/\bar{b}<<1$. In this case we can define the annulus (mean) aspect ratio, $\rho\equiv \bar{b}/\bar{a} =\tanh\bar{\nu}\leq 1$, as the ratio of the length of the major axis to the length of the minor axis and the annulus (mean) eccentricity as $e^2 \equiv 1-\rho^2=\sech^2\bar{\nu}\leq 1$. 
\noindent For a narrow confocal annulus, the expression of the local width is \cite{JPCM16}:
\vskip -8pt
\begin{equation}
\Delta w(\tau)=c\,\mathcal{Q}(\tau)\,\Delta\nu,
\label{width}
\end{equation}

\noindent where $\mathcal{Q}(\tau)$ is the elliptic scale factor defined by $\mathcal{Q}^2(\tau) \equiv \sinh^2\bar{\nu} \sin^2\tau+\cosh^2 \bar{\nu} \cos^2 \tau= \sinh^2\bar{\nu}+ \cos^2\tau=\cosh^2\bar{\nu} - \sin^2\tau=(\cosh2\bar{\nu} + \cos2\tau)/2$ that oscillates between $\sinh\bar{\nu}$ and $\cosh\bar{\nu}$ with a period $\pi$. In the small width approximation, $\Delta w_{max}<< \lambda_J$, where $\lambda_J$, called \Jos penetration length, gives a measure of the distance over which significant spatial variations of the \Jos phase occur, the \Jos phase of a CAJTJ does not depends on $\nu$ and the system becomes one-dimensional. The length of an elementary arc of the master ellipse is $ds=c \mathcal{Q}(\tau) d\tau$. Therefore, we introduce the non-linear curvilinear coordinate $s(\tau)=c \int_{0}^{\tau} \mathcal{Q}(\tau')d\tau'=c \cosh\bar{\nu}\,\texttt{E}(\tau,e^2)$, where $\texttt{E}(\tau,e^2)$ is the \textit{incomplete} elliptic integral of the second kind of modulus $e^2\leq1$. Accordingly, as $\tau$ changes by $2\pi$ then $s(\tau)$ increases by $L=\oint ds= 4c \cosh\bar{\nu}\,\texttt{E}(e^2)$ that is exactly the perimeter of the master ellipse. Here $\texttt{E}(e^2)\equiv \texttt{E}(\pi/2,e^2)$ is the {\it complete} elliptic integrals of the second kind of argument $e^2$. The mean perimeter of a narrow confocal annulus can also be expressed as $L=2\pi \overline{\Delta w}/ \Delta\nu$, where $\overline{\Delta w} \equiv (1/2\pi) {\int}_{-\pi}^\pi \Delta w(\tau)d\tau= (2/\pi) c \Delta \nu \cosh \bar{\nu}\,\texttt{E}(e^2)$ is the average annulus width. Then, the curvilinear coordinate $s(\tau)$ can be cast in a more compact form as $s(\tau)=L\,\texttt{E}(\tau,e^2)/4\,\texttt{E}(e^2)$; being $\texttt{E}(\tau,0)=\tau$ and $\texttt{E}(0)=\pi/2$, for a thin circular ring with mean radius $\bar{r}=L/2\pi$, it would be $s(\tau)=s(\theta)=\bar{r} \theta$. Furthermore, the area of a narrow confocal annulus is $A=\pi c^2 \cosh2\bar{\nu}\Delta\nu=\pi c \Delta w_{min} \cosh2\bar{\nu}/\sinh\bar{\nu}$. 



\vskip 5pt
\noindent It has been recently derived that the $\nu$-independent \Jos phase, $\phi(\tau,\hat{t})$, of a one-dimensional CAJTJ with a uniform critical current density, $J_c$, in the presence of a spatially homogeneous barrier-parallel magnetic field, ${\bf H}$, of arbitrary orientation, $\bar{\theta}$, relative to the $Y$-axis, obeys a modified and perturbed sine-Gordon equation with a space dependent effective Josephson penetration length inversely proportional to the local junction width \cite{JLTP16b}:
\vskip -8pt
\begin{equation}
 \left[\frac{\lambda_J}{c\,\mathcal{Q}(\tau)}\right]^2 \left(1+\beta\frac{\partial}{\partial \hat{t}}\right) \phi_{\tau\tau} - \phi_{\hat{t}\hat{t}}-\sin \phi =\alpha \phi_{\hat{t}} - \gamma(\tau) + F_h(\tau),
\label{psge}
\end{equation}

\noindent where $\hat{t}$ is the time normalized to the inverse of the so-called (maximum) plasma frequency, $\omega_p$. The subscripts on $\phi$ are a shorthand for derivative with respect to the corresponding variable. Furthermore, $\gamma(\tau)=J_Z(\tau)/J_c$ is the local normalized density of the bias current and 
\vskip -8pt
\begin{equation}
F_h(\tau)\equiv h\Delta \frac{\cos\bar{\theta}\cosh\bar{\nu} \sin\tau-\sin\bar{\theta}\sinh\bar{\nu}\cos\tau }{\mathcal{Q}^2(\tau)}
\label{Fh}
\end{equation}
\noindent is an additional forcing term proportional to the applied magnetic field; $h\equiv H/J_c c$ is the normalized field strength for treating long CAJTJs and $\Delta$ is a geometrical factor which has been referred to as the coupling between the external field and the flux density of the junction \cite{gronbech}. As usual, the $\alpha$ and $\beta$ terms in Eq.(\ref{psge}) account for, respectively, the quasi-particle shunt loss and the surface losses in the superconducting electrodes. Eq.(\ref{psge}) can be classified as a perturbed and modified sine-Gordon equation in which the perturbations are given by the system dissipation and driving fields, while the modification is represented by an effective local $\pi$-periodic \Jos penetration length, $\Lambda_J(\tau)\equiv \lambda_J/Q(\tau)= c \lambda_J \Delta \nu /\Delta\!W(\tau)$, inversely proportional to the annulus width. It is worth to point out that this $\Lambda_J$ variation stems from the variable junction width and cannot be modeled in terms of a spatially varying $\lambda_J$ in a uniform-width JTL as treated in Refs.(\cite{sakai},\cite{petras}); however, in the time independent case, it happens to be equivalent to a change in the $J_c$ of a uniform-width JTL \cite{semerdzhieva}. As the annulus aspect ratio approaches unity, the factor $c \mathcal{Q}$ tends to the ring radius and Eq.(\ref{psge}) reduces to the well-known perturbed sine-Gordon equation of a circular annular JTLs \cite{gronbech}. We stress that, for CAJTJs in a uniform in-plane magnetic field, the component of the applied magnetic field normal to the junction perimeter varies very smoothly, guaranteeing an accurate modeling at variance with other proposed geometries for a Josephson vortex qubit based on the $\delta$-like behavior of the normal field or of the local critical current density \cite{goldobin01,kemp02,kim11}. 
\vskip 14pt
\noindent As already said, when cooling an annular JTL below its critical temperature one or more magnetic flux quanta may be spontaneously trapped in its doubly connected electrodes; their trapping probability is known to increase with the speed of the normal-to-superconducting transition \cite{PRB09}. The algebraic sum of the flux quanta trapped in each electrode is an integer number $n$, called the winding number, counting the number of Josephson vortices (fluxons) trapped in the \jun barrier; also the spontaneous fluxon trapping process follows a statistical law \cite{PRB06}. In the absence of a symmetry-breaking external magnetic field the likelihoods to trap a fluxon or an antifluxon are equal \cite{PRB08}. Once trapped the fluxons can
never disappear and only fluxon-antifluxon pairs can be nucleated. To take into account the number of trapped fluxon, Eq.(\ref{psge}) is supplemented by periodic boundary conditions \cite{PRB96}:
\vskip -8pt
\begin{subequations}
\begin{eqnarray} \label{peri1}
\phi(\tau+2\pi,\hat{t})=\phi(\tau,\hat{t})+ 2\pi n,\\
\phi_\tau(\tau+2\pi,\hat{t})=\phi_\tau(\tau,\hat{t}).
\label{peri2}
\end{eqnarray}
\end{subequations}
\vskip -4pt

\subsection{ Single fluxon energy}

\noindent In the absence of dissipative and driving forces, the simplest topologically stable dynamic solution to Eq.(\ref{psge}) on an infinite line, in a first approximation, is a $2\pi$-kink (fluxon) centered at a time-dependent coordinate $s_0(\hat{t})$, namely, $\tilde{\phi}(\tau,\hat{t})= 4 \arctan \exp \left\{\wp[s(\tau)-s_0(\hat{t})]/\lambda_J \right\}$, where $\wp=\pm1$ is the topological charge, i.e., the fluxon polarity \cite{scott}. Indeed, the phase profile:
\vskip -8pt
\begin{equation}
\tilde{\phi}(\tau,\hat{t})= 4 \arctan \exp \left\{ \wp \left[ \frac{L\,\texttt{E}(\tau,e^2)}{\lambda_J\,\texttt{E}(e^2)} - \frac{s_0(\hat{t})}{\lambda_J} \right]\right\},
\nonumber
\end{equation}

\noindent satisfies Eq.(\ref{psge}) with damping and driving terms dropped, provided that \cite{JPCM16} the annular junction is long enough on the kink scale, $c/{\lambda_J} >>1/\sinh^3\bar{\nu}$, and both the normalized (tangential) fluxon speed, $\hat{u}\equiv d({s}_0/{\lambda_J})/d\hat{t}$, and acceleration, $\hat{a}\equiv d\hat{u}/d\hat{t}$, are (in moduli) much less than unity (non-relativistic limit). 

\noindent The Lagrangian and Hamiltonian densities associated with Eq.(\ref{psge}) have been derived in Ref.\cite{JPCM16}. By assuming that the annulus is long enough so that the left and right tails of the fluxon do not interact, it was found that in the absence of external forces the energy of a non-relativistic fluxon, $\hat{E}= \hat{K}+\hat{U}_w$, is conserved. The circumflex accents denotes normalized quantities. $\hat{E}$ is normalized to the characteristic energy, $\mathcal{E}=\Phi_0 J_c \lambda_J c \Delta \nu/2\pi$. Both the kinetic energy, $\hat{K}(\tau_0)\approx 4 \mathcal{Q}(\tau_0) \hat{u}^2$, and the intrinsic potential energy, $\hat{U}_w(\tau_0) \approx 8 \mathcal{Q}(\tau_0)$, are position dependent through the scale factor $\mathcal{Q}$ - see Eq.(\ref{width}). This is consistent with the relativistic expression $\hat{E}=\hat{m}(\tau_0)/\sqrt {1-\hat{u}^2(\tau_0)}$ reported by Nappi and Pagano \cite{nappipagano}, provided that we introduce the position dependent rest mass $\hat{m}(\tau_0)=8 \mathcal{Q}(\tau_0)$. Note that the energy of a CAJTJ containing one static vortex is ${m}(\pm \pi/2)=8\mathcal{E}$, with $c \mathcal{Q}(\pm \pi/2) \Delta \nu$ being the smallest annulus width.

\begin{figure}[tb]
\centering
\includegraphics[height=6cm]{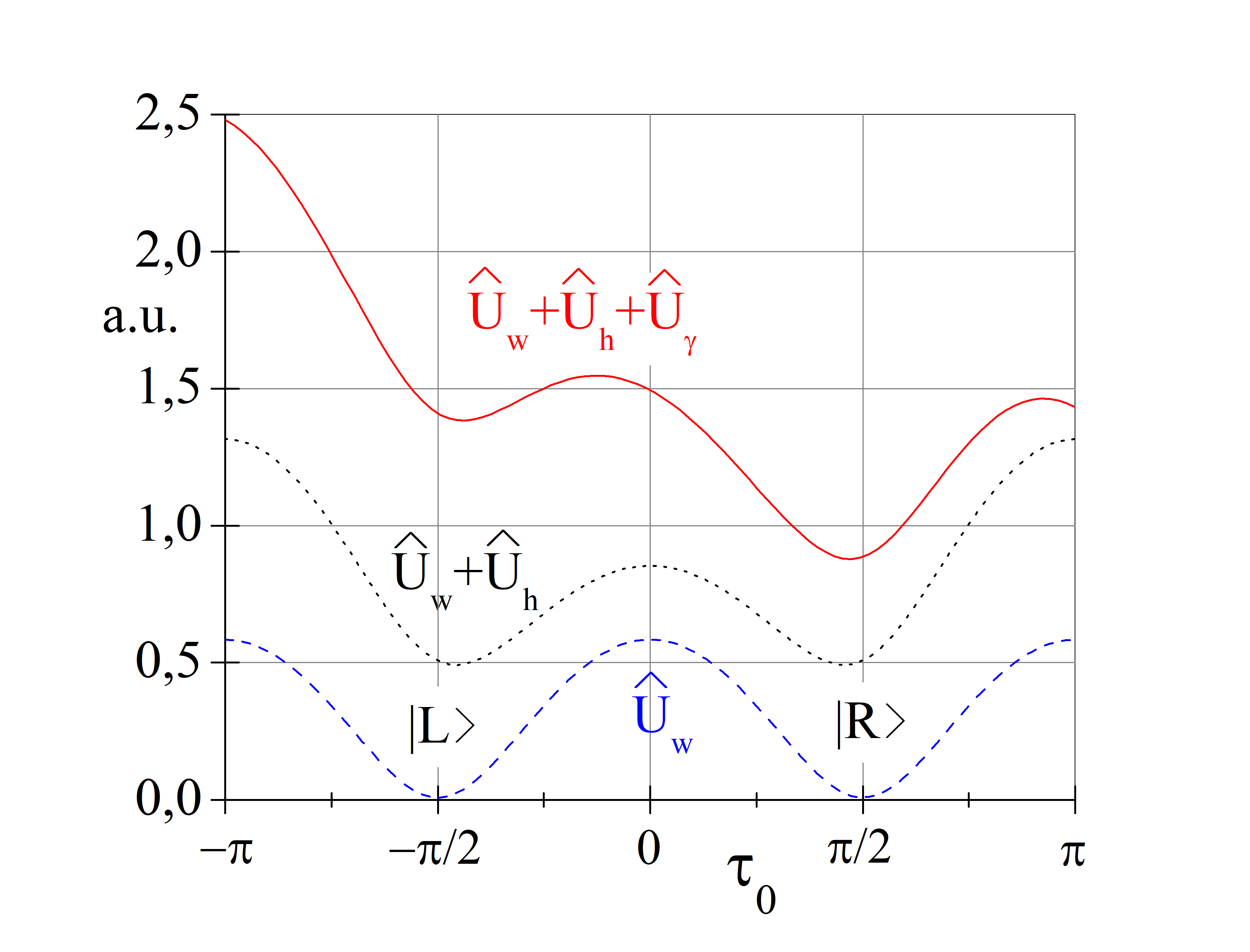}
\caption{(Color online) Schematic representation of the fluxon one-dimensional potential in different conditions. The dashed line refers to the intrinsic width-induced potential $\hat{U}_w$ with two minima at $\tau=\pm \pi/2$ coincident with the degenerate states $|R\rangle$ and $|L\rangle$; the dotted line corresponds to the symmetric double-well potential $\hat{U}_w+\hat{U}_h$ in the presence of a uniform magnetic field perpendicular to the long annulus diameter; the solid line show the potential $\hat{U}_w+\hat{U}_h+\hat{U}_\gamma$ in the most generic case of applied magnetic field and bias current. The three potentials are shifted by arbitrary vertical offsets.}
\label{potentials}
\end{figure}

\medskip
The potential $\hat{U}_w$, shown by the dashed curve in Figure~\ref{potentials}, expresses a $\pi$-periodic potential energy function uniquely determined by the CAJTJ ellipticity, $e^2\equiv \sech^2\bar{\nu}$. The potential wells are located at $\tau_0=\pm \pi/2$, where the annulus width is smallest. The left $|L\rangle$ and right $|R\rangle$ wells of the potential constitute stable classical states for the vortex with degenerate ground state energy. Considering that $\sinh\bar{\nu}\leq \mathcal{Q}(\tau) \leq \cosh\bar{\nu}$, the potential wells are separated by an energy barrier proportional to the exponential of $\bar{\nu}$. For a CAJTJ of moderate eccentricity (corresponding to $\rho\geq 0.5$), $\mathcal{Q}(\tau)$ can be approximated by its truncated Fourier expansion, $\mathcal{Q}(\tau)\approx (2/\pi)\cosh\bar{\nu} \,\texttt{E}(e^2) + \cos 2\tau/2\sqrt{2\cosh\,2\bar{\nu}}$, and the unperturbed potential, $\hat{U}_w$, turns into a sinusoidal potential whose properties have been well investigated both in the thermal and quantum-mechanical regimes \cite{wallraff00,wallraff03}. We stress that $\hat{U}_w$ is an intrinsic potential, i.e., it occurs in the absence of an applied magnetic field. However, it differs from the sinusoidal potential induced by a small uniform field applied to a circular annular JTL \cite{PRB98} in several aspects: i) $\hat{U}_w$ has an halved periodicity, i.e., there are two minima and two maxima for every round trip; ii) $\hat{U}_w$ is proportional to $\phi_\tau^2$ and so is independent on the fluxon polarity, $\wp$, while a magnetic potential complies with the fluxon polarity; iii) by squashing the annulus the relative inter-well barrier height can be made arbitrarily large, albeit limited by the resolution of the lithographic processes in the reproducing the annulus narrowest region.


\noindent In the general case with applied magnetic field and bias current, the one-dimensional potential energy experienced by the fluxon is made up by the sum of three contributions:
\vskip -8pt
\begin{equation}
\hat{U}(\tau_0)=\hat{U}_w(\tau_0)+\hat{U}_h(\tau_0)+\hat{U}_\gamma(\tau_0).
\label{Utot}
\end{equation}

\noindent $\hat{U}_h(\tau_0)\approx 2\pi \wp (\lambda_J/c) u_h(\tau_0)$ is the $2\pi$-periodic magnetic potential, where:
\vskip -8pt
\begin{equation}
u_h(\tau)\equiv h \Delta \left(\sin\bar{\theta}\sinh\bar{\nu}\sin\tau +\cos\bar{\theta}\cosh\bar{\nu} \cos\tau \right).
\label{uh}
\end{equation}
\vskip -4pt

\noindent $u_h(\tau)$ is $\pi$-antiperiodic in $\tau$, i.e., $u_h(\tau+\pi) =-u_h(\tau)$, then it averages to zero over one period. Furthermore, $du_h /d\tau=F_h(\tau) \mathcal{Q}^2(\tau)$. For a \Jos ring, with $\tau$ replaced by $\theta$ and $\bar{\nu}\to \infty$, we recover the sinusoidal magnetic potential \cite{PRB97}, $\hat{U}_h(\theta)\propto \cos(\bar{\theta}-\theta)$. The dotted curve in Figure~\ref{potentials} shows how the fluxon potential changes when a (negative) perpendicular field ($\bar{\theta}=0$), is applied to the CAJTJ; the potential $\hat{U}_w(\tau_0)+\hat{U}_h(\tau_0)$ is still invariant under parity transformation $(\tau_0 \to -\tau_0)$ and develops into a field-controlled symmetric potential with finite walls and two spatially separated minima. Increasing further the magnetic field, eventually the minima coalesce and the perturbed potential becomes single-welled.

\noindent Furthermore, $\hat{U}_\gamma(\tau_0)\approx 2\pi \wp (\lambda_J/c) u_\gamma(\tau_0)$ is the current-induced potential; assuming a uniform current distribution $\gamma(\tau)=\gamma_0$, it is:

\vskip -8pt
\begin{equation}
u_\gamma(\tau)\equiv \frac{\gamma_0}{2}\left( \tau \cosh2\bar{\nu} + \frac{1}{2}\sin2\tau\right),
\label{ugamma}
\end{equation}
\vskip -4pt
\noindent such that $du_\gamma /d\tau=\gamma_0 \mathcal{Q}^2(\tau)$. The solid line in Figure~\ref{potentials} shows the total potential when a bias current is feeding the CAJTJ. The resulting potential is qualitatively similar to the well-studied tilted washboard potential for the phase difference of a small JTJ biased below its critical current \cite{fulton74}; the only difference is that in our case the degree of freedom is the space, rather than the Josephson phase difference. Indeed, the potential profile can be tilted either to the left or to the right depending on the polarity of the bias current, $\gamma_0$. The inclination is proportional to the Lorentz force acting on the vortex which is induced by the bias current applied to the junction. The smallest tilt that allows the vortex to escape from a well defines the so-called depinning current, $\gamma_d$.

\section{ The numerical simulations}
 
\noindent In this section we numerically investigate the static and dynamic properties of a long and narrow CAJTJ in the presence of an external in-plane magnetic field applied along one of its symmetry axes. The commercial finite element simulation package COMSOL MULTIPHYSICS (www.comsol.com) was used to numerically solve Eq.(\ref{psge}) subjected to the cyclic boundary conditions in Eqs.(\ref{peri1}) and (\ref{peri2}) for several values of the winding number $n$. We set the damping coefficients $\alpha=0.05$ (weakly underdamped limit) and $\beta=0$, while keeping the current distribution uniform, i.e., $\gamma(\tau)= \gamma_0$. In addition, the field coupling constant, $\Delta$, was set equal to $1$. In order to compare the numerical results with the experimental findings presented in the next Section, we set the annulus aspect ratio to $\rho=1/2$ - as in Figure~\ref{ConfAnn} - corresponding to $e^2=0.75$ and $\bar{\nu}\approx 0.549$, such that the largest CAJTJ width is twice its smallest one (in fact, $\Delta w_{min}/\Delta w_{max}=\tanh \bar{\nu}=\rho$). Furthermore the normalized length was set to $\ell=L/\lambda_J=10\pi$; then, the (smooth) variation of the annulus width occurs over a length, $L/4=2.5 \pi \lambda_J \approx 8 \lambda_J$, quite large compared to the fluxon size.


\subsection{ The magnetic diffraction patterns: $n=0$}

\begin{figure}[t]
\centering
\subfigure[ ]{\includegraphics[height=5cm,width=7cm]{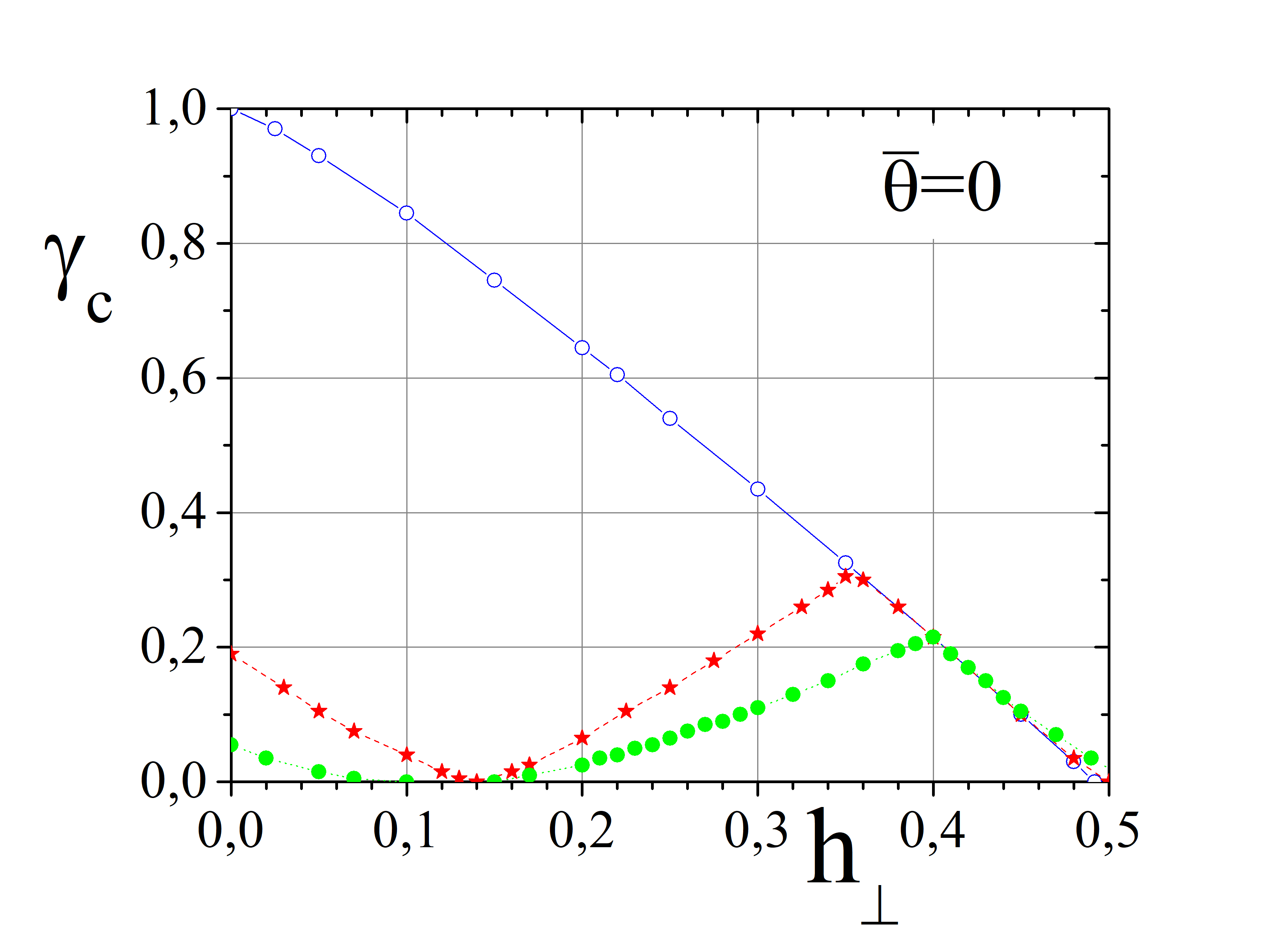}}
\subfigure[ ]{\includegraphics[height=5cm,width=7cm]{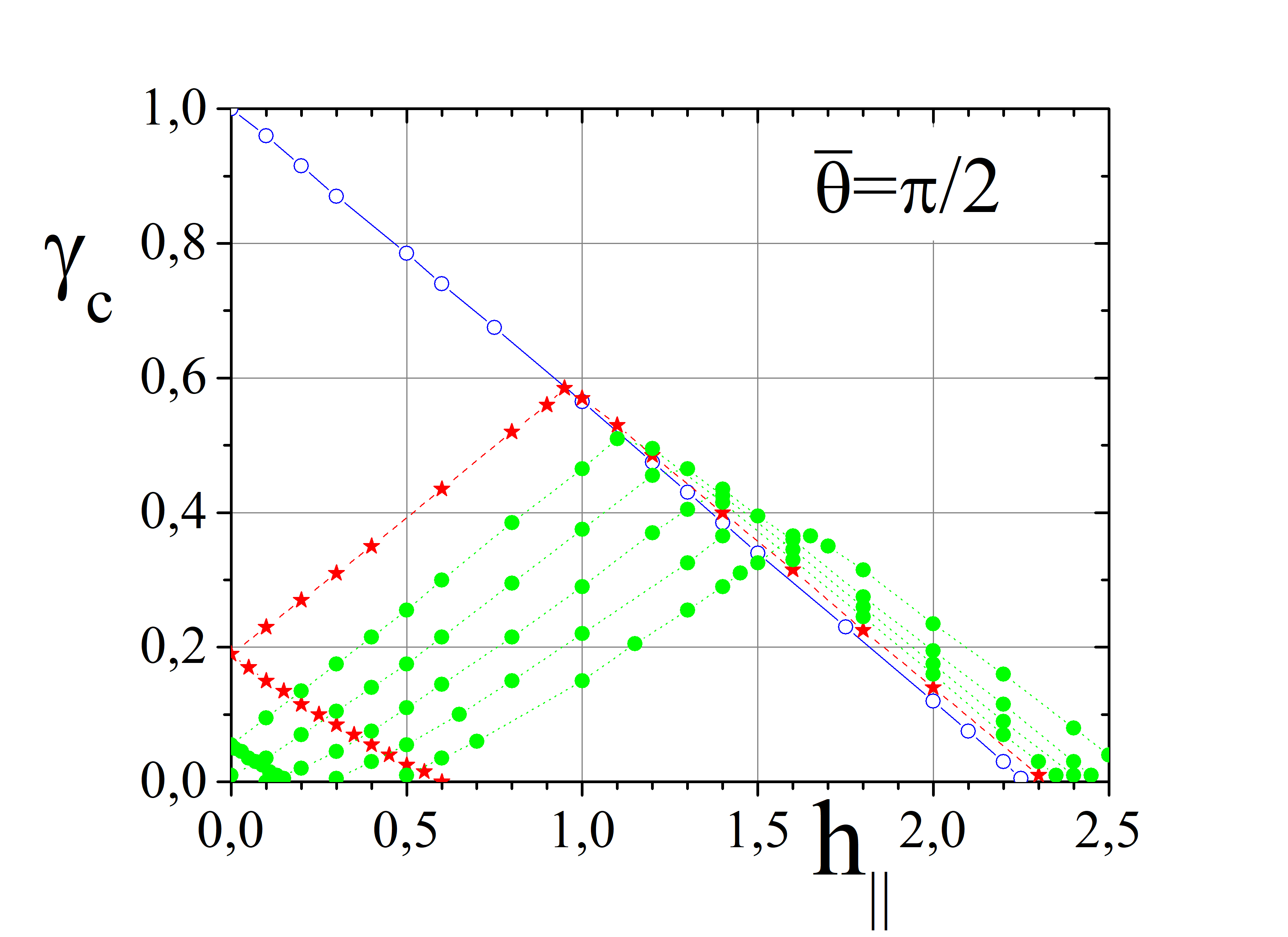}}
\caption{(Color online) Numerically computed magnetic diffraction patterns, $\gamma_c(h)$, of a one-dimensional CAJTJ with $\rho=0.5$, $\ell=10\pi$ and $n=0$ for two values of the in-plane field orientation, $\bar{\theta}$,: (a) $\bar{\theta}=0$, (b) $\bar{\theta}=\pi/2$. The magnetic fields are normalized to $J_c c$.}
\label{MDP}
\end{figure}

\noindent To begin with, numerical integrations of Eq.(\ref{psge}) have been carried out in the stationary, i.e., time-independent, state ($\phi_{\hat{t}}=0$) to derive the magnetic diffraction pattern (MDP) of the critical current of the CAJTJs. Specifically, we have numerically computed the maximum (or critical) value, $\gamma_c=I_c(H)/I_c(0)$, of the normalized zero-voltage current versus the normalized field amplitude, $h=H/J_c c$, in the case of no trapped fluxons ($n=0$). We considered two orthogonal orientations of the in-plane magnetic field relative to the annulus major diameter: a field $h_{\bot}$ perpendicular to the major axis corresponds to a field orientation $\bar{\theta}=0$ in the magnetic forcing term $F_h$ defined in Eq.(\ref{Fh}), vice versa for $\bar{\theta}=\pi/2$ the field is parallel to the major diameter and will be named $h_{\parallel}$. These coordinate-system-independent notations will turn out to be useful in the next section where we discuss the experimental results of CAJTJs whose foci lie either on the $X$-axis - as in Figure~\ref{ConfAnn} - or, by means of a $90^o$ rotation, on the $Y$-axis in the presence of a magnetic field applied along the $Y$-direction.


\medskip
\noindent The MDPs of electrically small ($\ell<<1$) and intermediate-length ($\ell=4\pi$) CAJTJs with $\rho=0.5$ have been reported in, respectively, Ref.\cite{JLTP16a} and Ref.\cite{JPCM16}. At variance with any previously considered long JTJ, the zero-field critical current was found to be multiple-valued due to the existence of static fluxon(s) and antifluxon(s) constrained either in the same width-induced potential well or in diametrically opposed wells until the Lorentz force associated with the bias current is strong enough to start their motion. In the pioneering paper by Owen and Scalapino for linear constant-width long JTJs \cite{owen}, the multiple-valued $\gamma_c$, corresponding to different configurations of the \Jos phase inside the barrier, were only observed in the presence of a magnetic field; this same behavior has been also confirmed in circular annular JTLs \cite{PRB96}. In Figures~\ref{MDP}(a) and (b) we show the numerically computed MDPs for a $10\pi$-long CAJTJ with $\rho=0.5$. Since, as far as $n=0$, it is $\gamma_c(-h)=\gamma_c(h)$, here we only show the dependence for positive field values.


\medskip
\noindent In order to trace the different lobes of the MDP, it is crucial to start the numerical integration with a proper initial phase profile, $\phi(\tau,0)$, compatible with the chosen winding number. In the figures we plot the solutions corresponding to the principal phase configurations. The mail lobe of the MDPs, denoted by open circles, was obtained by starting the integration with a spatially uniform phase profile and shows a linear decrease of the critical current with the external field; indeed, this feature, common to all long JTJs, can be erroneously interpreted as the signature of the full expulsion of the magnetic field from the junction interior (Meissner effect) that is not achievable in curved junctions \cite{SUST15}. The field value where the main lobe vanishes is called the (first) critical field, $h_{c1}$; we note that the critical field is smaller for $\bar{\theta}=0$, that is, as expected, the response to the external field is stronger when the field is perpendicular to the longest annulus diameter.

\medskip
\noindent The $\gamma_c$ values obtained with an initial phase configuration containing one fluxon-antifluxon ($F\bar{F}$) pair are identified by stars in the MDP plots. Initially, in the absence of a magnetic field, $F$ and $\bar{F}$ each has to be in its own potential well, otherwise they annihilate. As we increase the field, we observe different behaviors depending on the field direction. In the perpendicular field, $h_{\bot}$, that does not break the symmetry of the potential, no matters which particle is in which well and the critical current decreases to zero; upon increasing the field further, both particles fall in the same well and are kept apart by the magnetic force which prevents their annihilation. Conversely, in a small parallel field, $h_{\parallel}$, the modulation depends on the initial positions of the particles: $\gamma_c$ increases when the fluxon is in the right well and the antifluxon in the left well, while it decreases in the opposite case. The full circles correspond to higher lobes resulting from an initial phase profile containing more than one $F\bar{F}$ pair. For $\bar{\theta}=0$ we could only find solutions corresponding to one and two static pairs. On the contrary, by rotating the field by $90^o$, solutions with up to ten pairs could be easily found; in Fig~\ref{MDP}(b) we show the numerical data up to six $F\bar{F}$ pairs. 



\subsection{ Single fluxon statics: $n=+1$}

\noindent In principle, the static properties of a fluxon trapped in a one-dimensional CAJTJ, could be disclosed by minimizing the potential in Eq.(\ref{Utot}), i.e., by finding the roots of $d\hat{U}/d\tau_0$ and then selecting the stable $\tau_0$-positions. However, this process would provide approximate results when the external potentials, $\hat{U}_h$ and $\hat{U}_\gamma$ cannot be considered as small perturbations and, even worse, when the CAJTJ is not very long. Furthermore, the collective coordinate describing the motion of topological solutions of the sine-Gordon equation was introduced with the assumption that the fluxon is a rigid body, whose shape does not change when it moves. This condition is not fulfilled in varying-width JTls in which the fluxon has a position-dependent inertial mass. For the reasons above, the potential in Eq.(\ref{Utot}) is useful just for a qualitative understanding and it is mandatory to resort to numerical analysis. The manipulation of the vortex states, that are important with respect to the possible realization of a vortex qubit, are numerically analyzed in this subsection.

\begin{figure}[b]
\centering
\subfigure[ ]{\includegraphics[height=5cm,width=7cm]{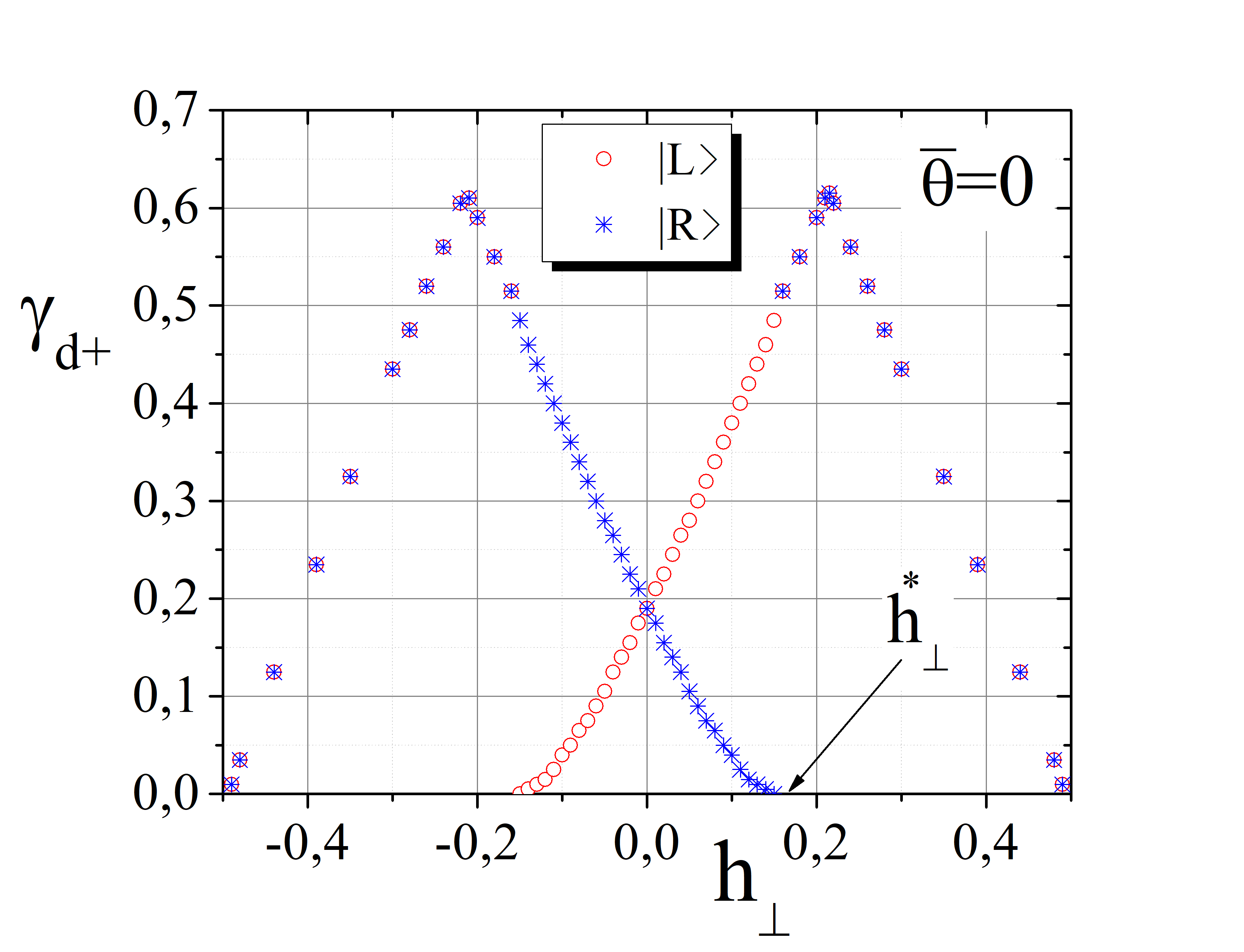}}
\subfigure[ ]{\includegraphics[height=5cm,width=7cm]{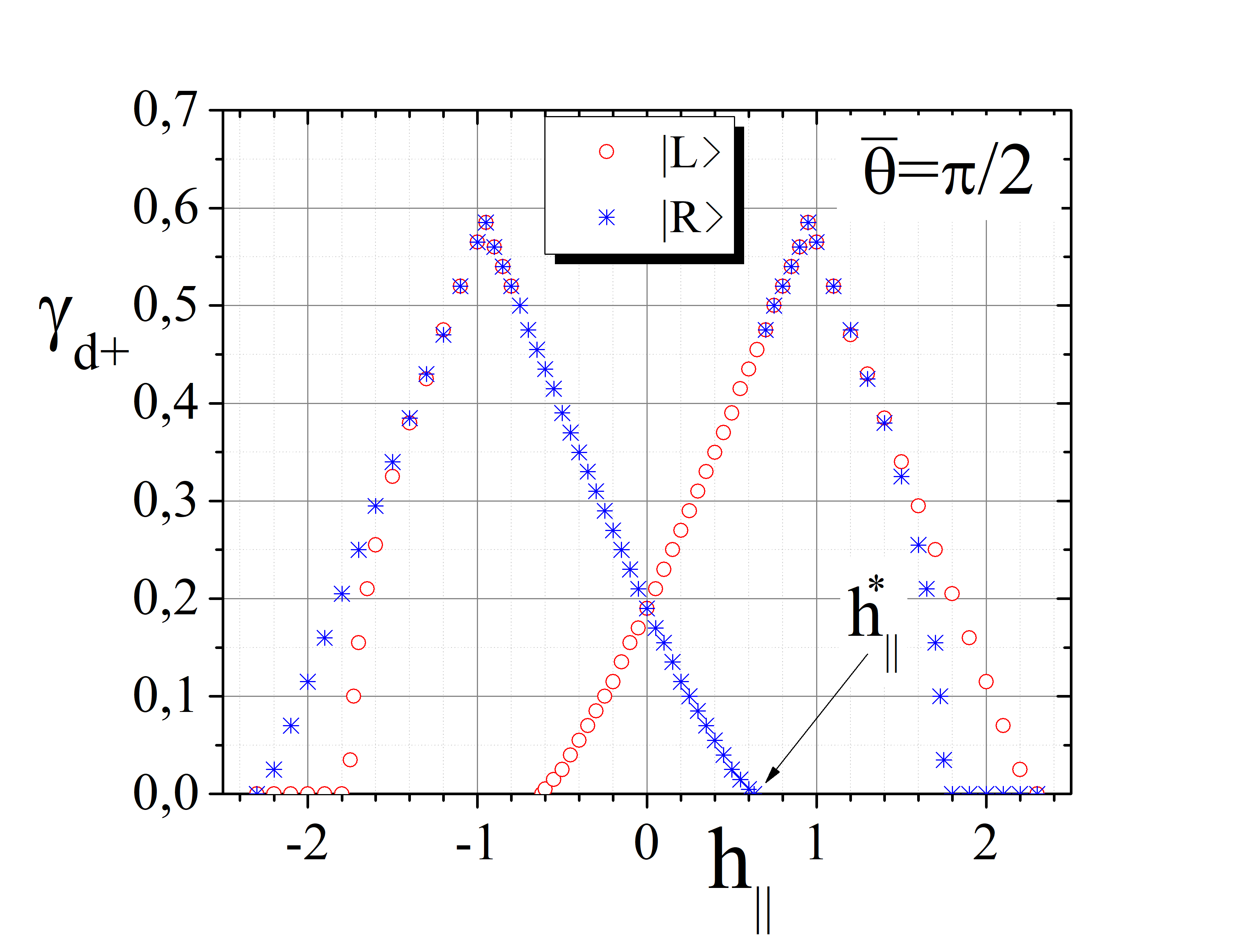}}
\caption{(Color online) Numerically computed field dependence of the positive fluxon depinning currents of the $|L\rangle$ (open circles for $\gamma_{d+}^{L}$) and $|R\rangle$ (stars for $\gamma_{d+}^{R}$) states for two values of the in-plane field orientation, $\bar{\theta}$,: (a)  $\bar{\theta}=0$, (b) $\bar{\theta}=\pi/2$. The magnetic fields are normalized to $J_c c$.}
\label{depinning}
\end{figure}
 
\medskip
\noindent  A static fluxon centered either in left ($\tau_0=-\pi/2$) or right well ($\tau=\pi/2$) was chosen for the system initial condition in Eq.(\ref{psge}) with $n=+1$. We first consider the case of an in-plane field, $h_{\bot}$, applied perpendicular to the longest annulus diameter, i.e., along the $Y$-axis ($\bar{\theta}=0$).  The numerical analysis showed that, for small $h_{\bot}$, the fluxon static positions in the equatorial points shift towards one of the polar points (depending on the field sign) until they merge for a (perpendicular) threshold field strength whose absolute value, $h_{\bot}^{*}\approx 0.16$, is well below the (first) perpendicular critical field. It follows that for $|h_{\bot}|\geq h_{\bot}^{*}$ the information about the vortex initial state is lost. For $|h_{\bot}|< h_{\bot}^{*}$, as we apply a bias current, the potential is tilted and at some point the fluxon is depinned from its original well and gets trapped in the opposite well which has an higher depinning current; this occurs because the intra-well barrier is much larger than the inter-well barrier \cite{carapella04}. The application and the later removal of the proper perpendicular field and bias current represent a viable procedure to prepare the vortex state; however, in the presence of a perpendicular field the state readout cannot be achieved by a current switch measurement.

\medskip
\noindent We now revert to the more interesting case of an in-plane field, $h_{\parallel}$, parallel to the longest annulus diameter ($\bar{\theta}=\pi/2$), whose potential, $\hat{U}_h$, is in phase with one of the wells of the intrinsic potential $\hat{U}_w$ and out of phase with the other one. Therefore, a sufficiently large parallel field, $h_{\parallel}^{*}$, will further deepen one well, while completely suppressing the other one. It means that any field value whose absolute value is larger than the parallel threshold field, $|h_{\parallel}|>h_{\parallel}^{*}\approx 0.62$, forces the fluxon in a given state - either $|L\rangle$ or $|R\rangle$ depending on the field polarity - without the need to apply a bias current. Once the qubit state has been prepared, the parallel field can be reduced or even removed. In addition, for $|h_{\parallel}|<h_{\parallel}^{*}$, the depinning currents for the two states are quite different and once depinned the fluxon has enough energy not to be re-trapped in the next well; it is, then, possible to discriminate between the two states by a current switch measurement. We are, of course, assuming that the losses are not so large to quickly dissipate the fluxon energy. 
\medskip
\noindent Our numerical findings on the fluxon static properties are summarized in the Figures~\ref{depinning}(a) and (b) reporting the field dependence of the positive depinning currents, $\gamma_{d+}^{L}$ (open circles) and $\gamma_{d+}^{R}$ (crosses), for a fluxon, respectively, either in the $|L\rangle$ or $|R\rangle$ initial state. Figures~\ref{depinning}(a) and (b) refer to a CAJTJ having $\rho=0.5$ in the presence of a, respectively, perpendicular and parallel in-plane magnetic field. We first note that the zero-field depinning currents are degenerate, $\gamma_{d+}^{L}(0)=\gamma_{d+}^{R}(0)$, and are an appreciable fraction of the zero-field critical current. As the magnetic field is turned on, it is seen that, the degeneration is removed.

\medskip
\noindent In the perpendicular field range $|h_{\bot}|<|h_{\bot}^*|$, the fluxon escaping from the well with the smaller depinning current is re-trapped in the other well which has a larger depinning current. Figure~\ref{depinning}(a) also shows that for $|h_{\bot}|\geq |h_{\bot}^*|$ the depinning currents abruptly become identical, $\gamma_{d+}^{L}(h_{\bot}) =\gamma_{d+}^{R}(h_{\bot})$. This occurs because the two wells have coalesced into a single well. For the negative depinning currents, $\gamma_{d-}^{L,R}$, it was found that $\gamma_{d-}(h_{\bot})= -\gamma_{d+}(-h_{\bot})$. Furthermore, a current inversion was found to correspond to an exchange of the $|L\rangle$ and $|R\rangle$ states, i.e., $\gamma_{d-}^{R} (h_{\bot})= -\gamma_{d+}^{L}(h_{\bot})$.

\medskip
\noindent As shown in Figure~\ref{depinning}(b), in the presence of an parallel field, $h_{\parallel}$, not only the depinning from the $|L\rangle$ and $|R\rangle$ states occurs at different bias currents in a quite large field range, $|h_{\parallel}|<h_{\parallel}^*$, but, once escaped, the fluxon is not re-trapped in the adjacent well. Therefore the measurement of the depinning current allows to localize the vortex in one of the two states. For the negative depinning currents, $\gamma_{d-}^{L,R}$, it was found that $\gamma_{d-}(h_{\parallel})= -\gamma_{d+}(h_{\parallel})$. It follows that the determination of the fluxon state can be as well accomplished through the measurement of a negative current switch. It might happen that for $|h_{\parallel}|>h_{\parallel}^*$ a small range of magnetic field exists for which $\gamma_{d+}^{L}(h_{\parallel})= \gamma_{d+}^{R}(h_{\parallel})$.


\noindent Summarizing, the fluxon state preparation can be reliably achieved by just applying a parallel field whose absolute value is larger than the parallel threshold field, $h_{\parallel}^*$, while the state read-out can be accomplished by a measurement of the depinning current in a smaller field, $|h_{\parallel}|<h_{\parallel}^*$.  Interestingly, if an antifluxon, rather than a fluxon, is trapped in the CAJTJ, the preparation and read-out procedure both work in the opposite way so that the final result is unchanged. It means that at the end of each successful trapping procedures, it will not be possible to determine the polarity of the spontaneously trapped fluxon. Plots qualitatively similar to those in Figures~\ref{depinning}(a)-(b), albeit with smaller depinning currents, have been obtained (but are not shown) for two trapped unipolar fluxons (though multiple spontaneous trappings are less likely to happen).


\subsection{ Single fluxon dynamics}

If a depinned fluxon has enough energy to escape all potential wells, it starts to travel around the annulus and a voltage jump from the static state ($V=0$) to the running state ($V\neq 0$) is detected. Generally speaking, the fluxon motion in current biased JTLs is manifested by a stable finite-voltage current branch in its current-voltage characteristic (IVC); in the absence of an external magnetic field this current singularity is called the (first) Zero-Field Step (ZFS1). In normalized units, the dc-current corresponds to the uniform forcing term $\gamma$ in Eq.(\ref{psge}), while the dc-voltage generated by the fluxon traveling around the annulus with revolution period $T$ is given by its spatio-temporal average speed $\bar{u}=\ell/T$; the asymptotic voltage of the resonance, $\bar{u}=1$, corresponds to an average speed equal to the Swihart velocity \cite{Swihart}, $\bar{c}$, which is the characteristic velocity of electromagnetic waves in JTLs. The tangential fluxon speed $\hat{u}$ increases (decreases) when it approaches a well (barrier) of the tilted periodic potential. This makes the fluxon dynamics in a CAJTJ very different from the constant speed motion in a uniform-width ring shaped JTJ. In fact, when a fluxon is accelerated other excitations such as the so-called plasma waves are radiated. Depending on the fluxon velocity, resonances may occur between the fluxon and the plasma waves corresponding to different wave numbers \cite{pedersen86}. Although the dispersion relation is not know in a CAJTJ, the strength of the resonance drastically depends on the waves amplitudes which, in turn, are strictly related on the system's dissipation and circumference as well as on the steepness of the potential energy difference. These resonances appear as regular fine-structures on the ZFS profile of samples with very low damping \cite{JAP93}.


\begin{figure}[t]
\centering
\includegraphics[height=6cm,width=8cm]{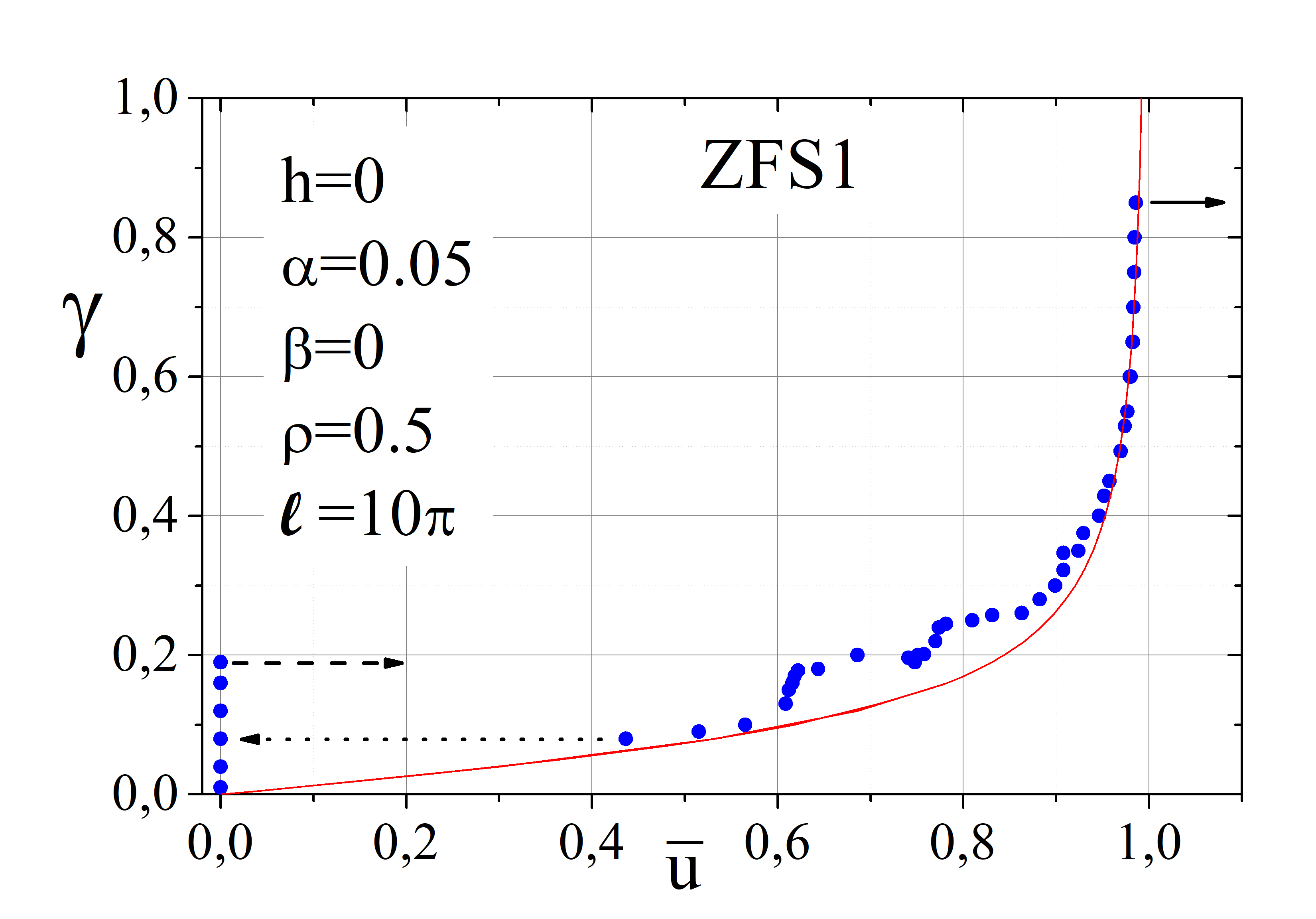}
\caption{(Color online) Numerically computed profile of the first zero-field step for a CAJTJ. Results are calculated integrating Eq.(\ref{psge}) with $\rho=0.5$, $l=10\pi$, $\alpha=0.05$, $\beta=0$, $h=0$, and $n=1$. The solid line is the perturbative model expectation $\gamma(\bar{u})= 4\alpha/\pi \sqrt{\bar{u}^{-2}-1}$ for a constant-width circular annular JTJ, i.e., for $\rho=1$.}
\label{ZFS1h0}
\end{figure}

\medskip
\noindent Figure~\ref{ZFS1h0} shows the numerically computed ZFS profile (i.e., $\gamma$ versus $\bar{u}$) of a CAJTJ having an aspect ratio $\rho=0.5$ and a normalized length $\ell=10\pi$. The dashed right pointing arrow at $\gamma=0.19$ indicates the depinning current already discussed in the previous paragraph, while the dotted left pointing arrow $\gamma=0.08$ denotes the re-trapping current, i.e., the minimum current at which the fluxon still moves along the system, not being trapped by the potential. The $\gamma-\bar{u}$ plot is quite smooth and only moderately departs from the perturbative model expectation $\gamma(\bar{u})= 4\alpha/\pi \sqrt{\bar{u}^{-2}-1}$ (solid line) valid for fluxon traveling in a flat potential and in the absence of collisions \cite{scott,JLTP16a}, that is, for $\rho=1$. The main discrepancy is observed for low bias currents, where the step profile is not smooth but shows some, not well resolved, fine structures \cite{JAP93} due to the resonance of the traveling fluxon with wavelets radiated by the fluxon itself subject to periodic accelerations and decelerations. The ZFS profile presents a premature switching point, indicated by the solid arrow at $\gamma=0.85$, due to the fluxon instability at high speed that prevents the fluxon from reaching relativistic speeds.

\section{ The measurements} 

\subsection{ The samples and the experimental setup}

\begin{figure}[t]
\centering
\includegraphics[width=6cm]{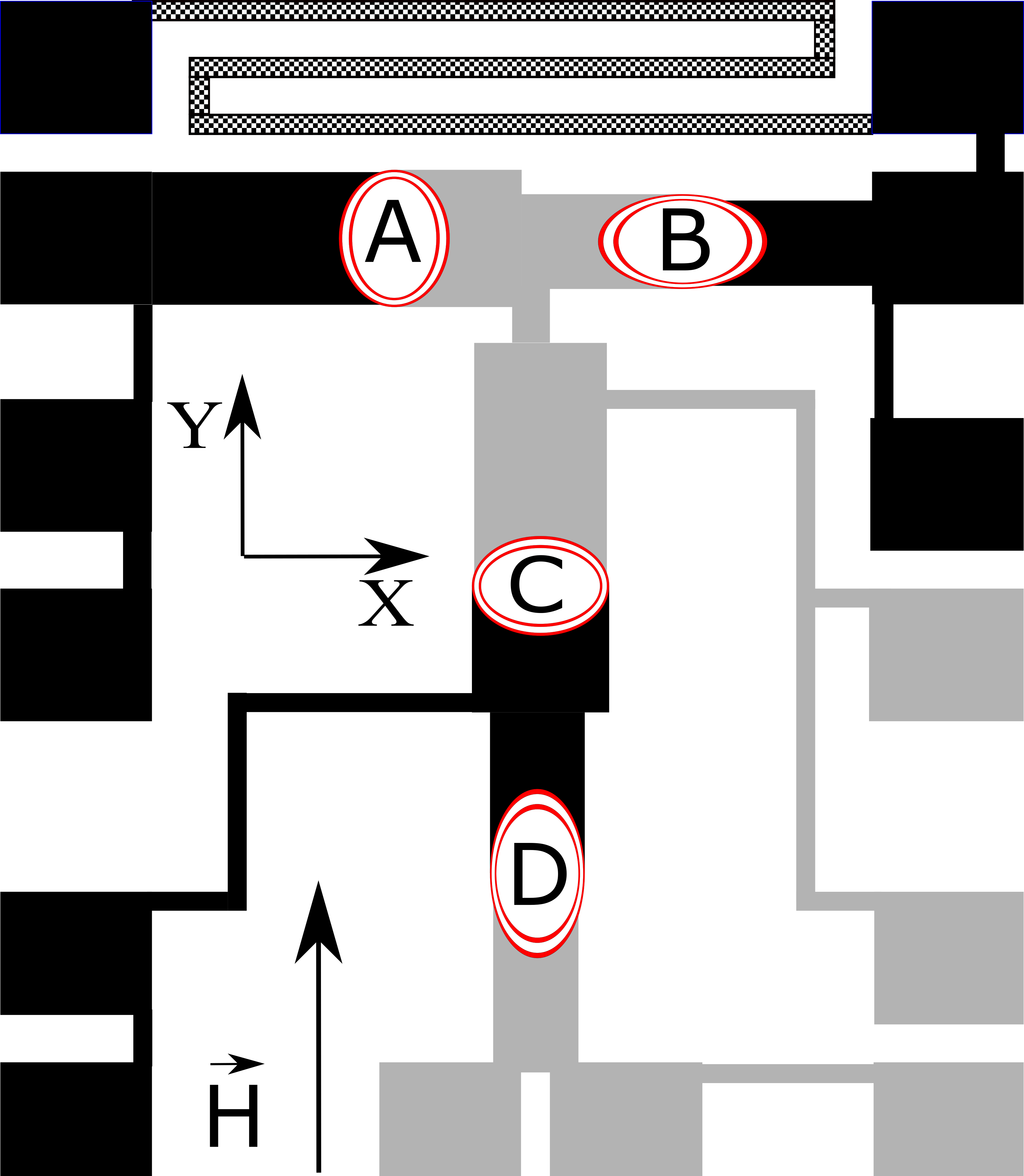}
\caption{(Color online) Layout of a $3\times4.2mm^2$ $Si$ chip each integrating four nominally identical CAJTJs; the top/wiring layer is in black, while the bottom layer is in gray. JJB and JJC have the longer principal axis along the $X$-axis, but differ with respect to the directions of the bias current. JJA and JJD have the longer principal axis along the $Y$-axis, but different bias current directions. The hatched meander-line strip on the top is a Mo resistive film used for a fast and reliable heating of the chip.}
\label{layout}
\end{figure}

\noindent Using the well known and reliable selective niobium etching and anodization process\cite{sneap} we have realized high-quality window-type $Nb/Al$-$AlOx/Nb$ CAJTJs. The details of the trilayer deposition and of the fabrication process can be found elsewhere \cite{VPK}. Four CAJTJs (named from JJA to JJD) are integrated on a $3\times4.2mm^2$ $Si$ chip all having the so called \textit{Lyngby-type} geometry\cite{davidson85} that refers to a specularly symmetric configuration in which the width of the current carrying electrodes matches one of the ellipse outer axis and the tunneling area is obtained by the superposition of two superconducting rings. The chip layout is schematized in Figure~\ref{layout}, where the top/wiring layer is shown in black and the bottom layer in gray. The hatched meander-line strip on the top is a Mo resistive film used for a fast and reliable heating of the chip; this resistive element has a nominal dc resistance of $100\,\Omega$ at LHe temperatures and, due to its good adhesion with the substrate, is very effective in dissipating heat to the chip. The junctions JJB and JJC, that we will call horizontal junctions, have the longer principal axis along the $X$-direction, but differ with respect to the direction of the bias current and the associated induced magnetic field (the so-called self-fields). On the contrary, the vertical junctions, JJA and JJD, have the foci on the $Y$-axis and again differ in the bias current direction. The reasons to have differently oriented CAJTJs are twofold: i) in our experimental set-up the barrier-parallel magnetic field can only be applied along the $Y$-axis; ii) generally speaking, the self-field in a long annular JTJ can be compensated by means of a magnetic field perpendicular to the direction of the bias current \cite{PRB96,SUST15}.

\medskip
\noindent All four CAJTJs on the chip had the same aspect ratio $\rho=0.5$ and annulus mean perimeter was $L=200\mu m$. The geometrical details of the CAJTJ tunneling area are listed in Table I. In designing the photo-lithographic mask which defines the area of the junctions, a 0,45 um under-etch occurring at the mask fabrication and subsequent junction definition process was taken into account; it means that on the mask the nominal barrier area is not delimited by two closely spaced confocal ellipses, but by one curve {\it parallel} to and inside the inner ellipse and another one {\it parallel} to and outside the outer ellipse. 

\begin{table*}[b]
	\centering
	 		\begin{tabular}{|c|c|c|c|c|c|c|c|c|c|c|c|}
		 \hline
     $\rho$ & $\bar{\nu}$ & $a_i$ & $b_i$ & $a_o$ & $b_o$ & $\Delta w_{min}$ & $\Delta w_{max}$ & c & A & L & $\Delta \nu$\\
		\hline
		  & & $\mu m$ & $\mu m$ & $\mu m$ & $\mu m$ & $\mu m$ & $\mu m$ & $\mu m$ & $\mu m^2$ & $\mu m$ &\\
		\hline
			 0.5 & 0.549 & 40.3 & 18.6 & 42.4 & 22.8 & 2.1 & 4.2 & 35.8 & 680 & 200 & 0.102\\
    \hline
    				\end{tabular}
		\caption{The geometrical details of the tunneling area of our CAJTJs. Refer to Figure~\ref{ConfAnn}. $\Delta \nu=\Delta w_{min}/c \sinh \bar{\nu}$.}
		 \label{table}
\end{table*}

\medskip
\noindent Our setup consisted of a cryoprobe inserted vertically in a commercial $LHe$ dewar. The chip with the CAJTJs is mounted on a Cu block enclosed in a vacuum-tight can immersed in the liquid He bath. The cryoprobe was magnetically shielded by means of two concentric long cylindrical $Pb$ cans and a cryoperm one; in addition, the measurements were carried out in an rf-shielded room. The external magnetic field could be applied both in the chip plane or in the orthogonal direction. The chip was positioned in the center of a long superconducting cylindrical solenoid whose axis was along the $Y$-direction (see Figure~\ref{layout}) to provide an in-plane magnetic field, either $H_{||}$ or $H_{\bot}$ depending on the junction orientation. The transverse magnetic field, $H_z$, was applied by means of a superconducting cylindrical coil with its axis oriented along the $Z$-direction. The field-to-current conversion factor was $3.9\, \mu$T/mA for the solenoid and $4.4\, \mu$T/mA for the coil. 

\medskip
\noindent 

\medskip
\noindent The critical current density of our samples was measured on electrically small cross-type \juns realized in the same wafer on different chips; at $T=4.2\, K$, we found $J_c\approx 2.2\,kA/cm^2$ corresponding to $\lambda_J\approx 5.9 \mu m$. Taking into account a $1.5\,\mu m$ wide idle region, it is $\lambda_J\approx 6.2 \mu m$ which provides a normalized length $\ell=L/\lambda_J\approx 32 \simeq 10 \pi$. We point out that the smallest curvature radius, $\bar{b}^2/\bar{a}= c \rho \sinh\,\bar{\nu}\approx 10.3 \mu m$ is larger than $\lambda_J$. A large number of samples were investigated whose high quality has been inferred by a measure of the their IVCs at $T=4.2\,K$. In fact, the subgap current $I_{sg}$ at $2\,mV$ was small compared to the current rise $\Delta I_{g}$ in the quasi-particle current at the gap voltage $V_{g}\approx 2.95\,mV$, typically $\Delta I_{g}>35I_{sg}$. In addition, all samples showed not only the zero-field critical current, $I_{c,0}$, but also the maximum critical current, $I_{c,max}$, considerably smaller than about the $70\%$ of the current jump at the gap voltage, $\Delta I_g$, typical of short $Nb/Al$-$AlOx$-$Al/Nb$ junctions. This is the first signature of a non-uniform bias current distribution and of the self-field effects \cite{SUST13a,SUST15}. More information about these two effects will be envisaged by analyzing the junctions magnetic diffraction patterns.

\subsection{ In-plane magnetic diffraction patterns} 

\begin{figure}[t]
\centering
\subfigure[ ]{\includegraphics[width=7cm]{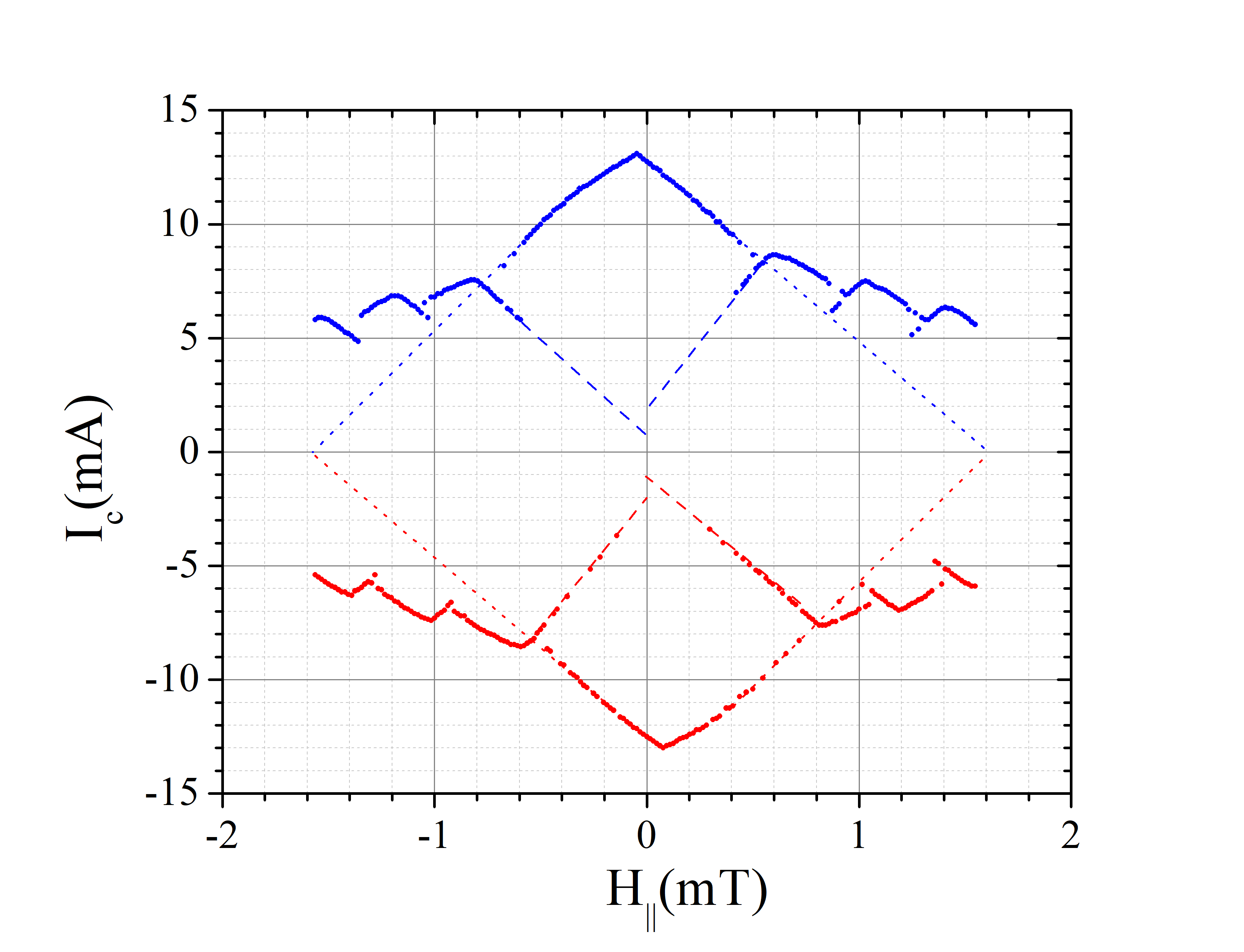}}
\subfigure[ ]{\includegraphics[width=7cm]{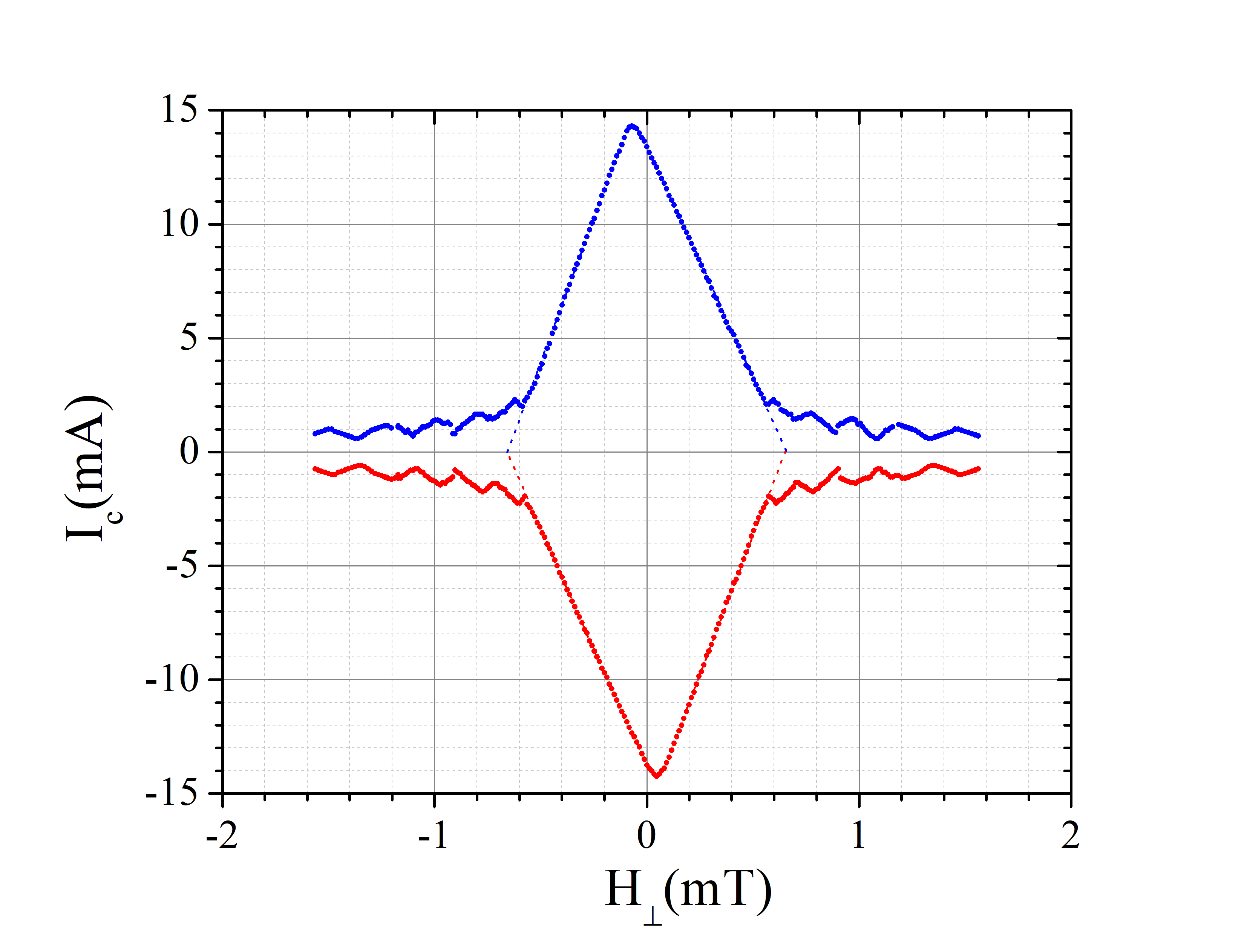}}
\subfigure[ ]{\includegraphics[width=7cm]{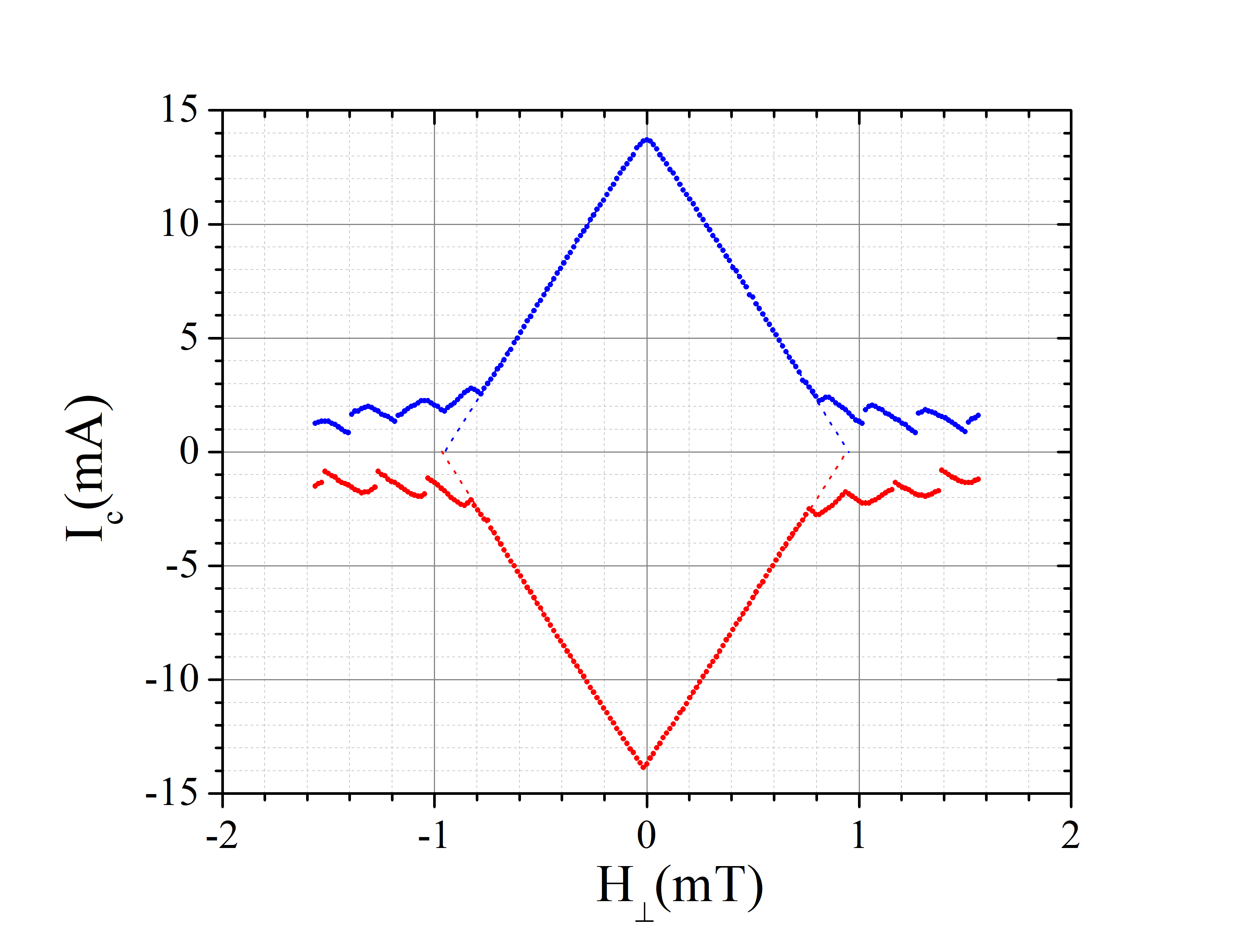}}
\subfigure[ ]{\includegraphics[width=7cm]{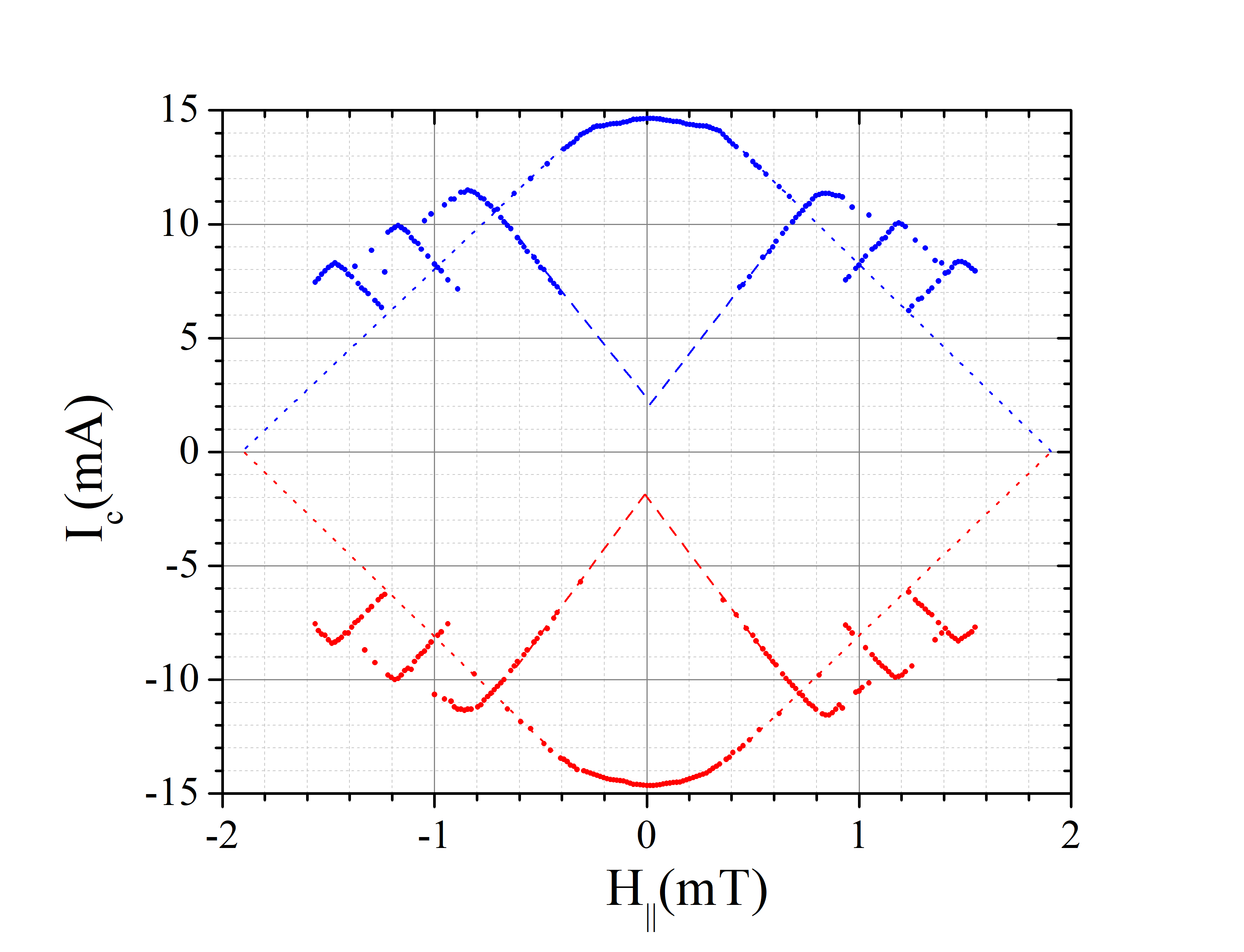}}
\caption{(Color online) Experimental magnetic diffraction patterns of the four nominally identical CAJTJs on a representative chip in the presence of an in-plane magnetic field applied along the $Y$-direction: (a) \jun A, (b) \jun B, (c) \jun C, (d) \jun D. The extrapolated dotted lines help to locate the critical fields.}
\label{par}
\end{figure}

On real devices, the measurements of maximum supercurrent against the external field often yield the envelop of the lobes, i.e., the current distribution switches automatically to the mode which for a given field carries the largest supercurrent. Sometimes, for a given applied field, multiple solutions are observed on a statistical basis by sweeping many times on the \jun IVC. 

\noindent Several chips were tested all made within the same fabrication run and they gave qualitatively similar results; the finding presented here pertain to just a representative one. Figures~\ref{par}(a)-(d) display the MDPs of the four junctions JJA to JJD on that chip ($\Delta I_g\approx 28\,mA$) with an in-plane magnetic field applied in the $Y$-direction; the (first) critical fields are obtained extrapolating to zero the MDP first lobe (see dotted lines). At a first glance, we observe that the MDPs of the horizontal junctions, JJB and JJC, for which the field is perpendicular to the longest annulus diameter are quite different from those of JJA and JJB whose longest diameter is parallel to the applied field. The main difference is the lack of multiple solutions for the horizontal junctions as compared to the pronounced overlapping lobes in the vertical junctions. This is the first experimental evidence of the width non-uniformity; in fact, for elliptic annular junctions of constant width, $I_c(H_{||})$ and $I_c(H_{\bot})$ simply scale with the inverse of the diameter perpendicular to the applied field \cite{SUST15}. We also observe that the MDPs of the topmost CAJTJs, JJA and JJB, are slightly skewed: this is a well known effects occurring in long annular junctions whose bias current is perpendicular to direction of the applied magnetic field \cite{PRB96,SUST15}. The comparison of the experimental MDPs with those expected - see Figures~\ref{MDP}(a) and (b) - highlights several common features such as the parallel critical fields ($> 1.5\,mT$) being larger than the orthogonal ones ($< 0.8\,mT$) and the large amplitude of the secondary lobes in the presence of a parallel field. The main discrepancy is the absence of zero-field double solutions in the experimental MDPs; however, at least for the horizontal CAJTJs, the linear extrapolations to zero field of the first positive and negative lobes, as indicated by the dotted line, converge to finite current values.

\subsection{ Transverse magnetic diffraction patterns}

\begin{figure}[t]
\centering
\subfigure[ ]{\includegraphics[width=7cm]{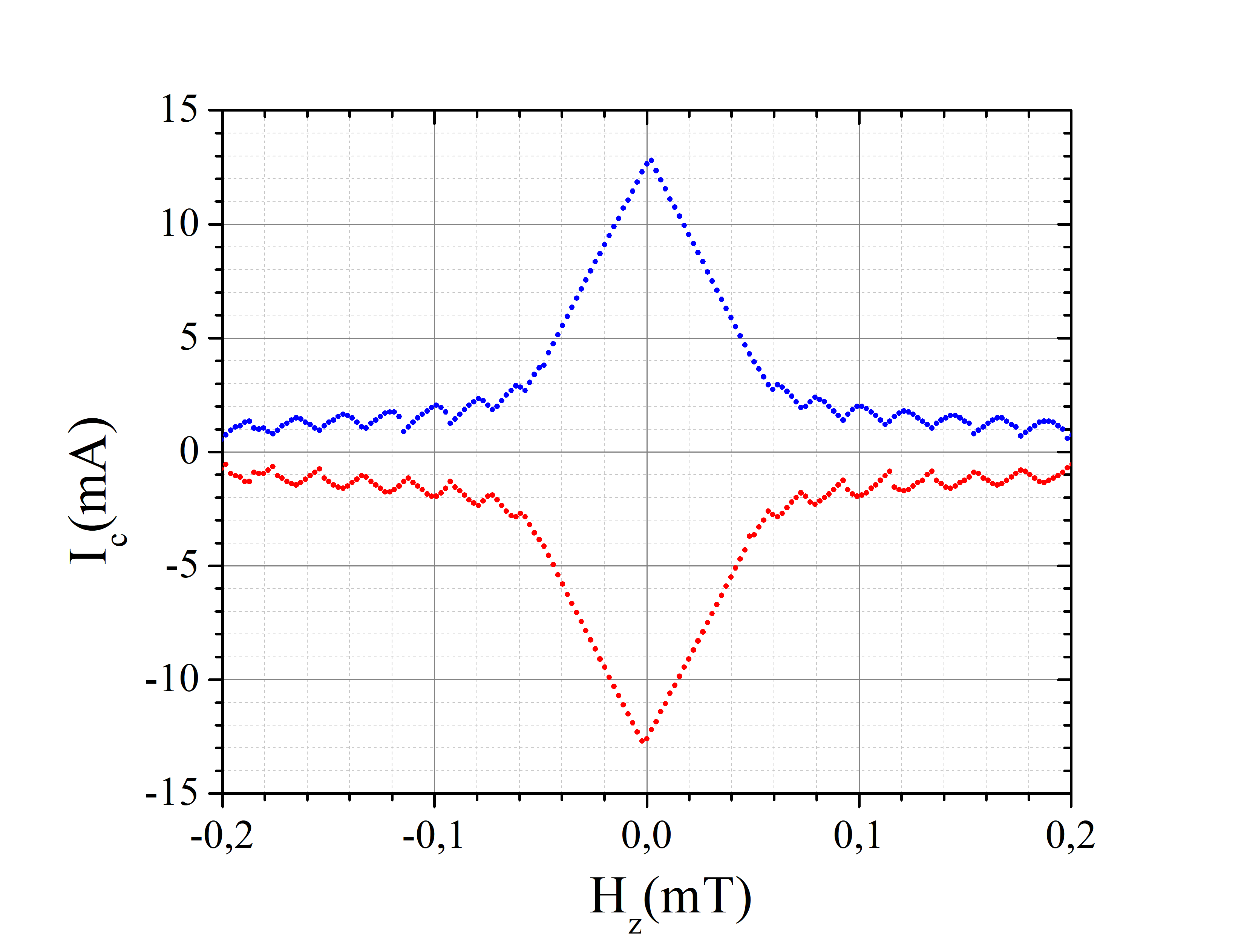}}
\subfigure[ ]{\includegraphics[width=7cm]{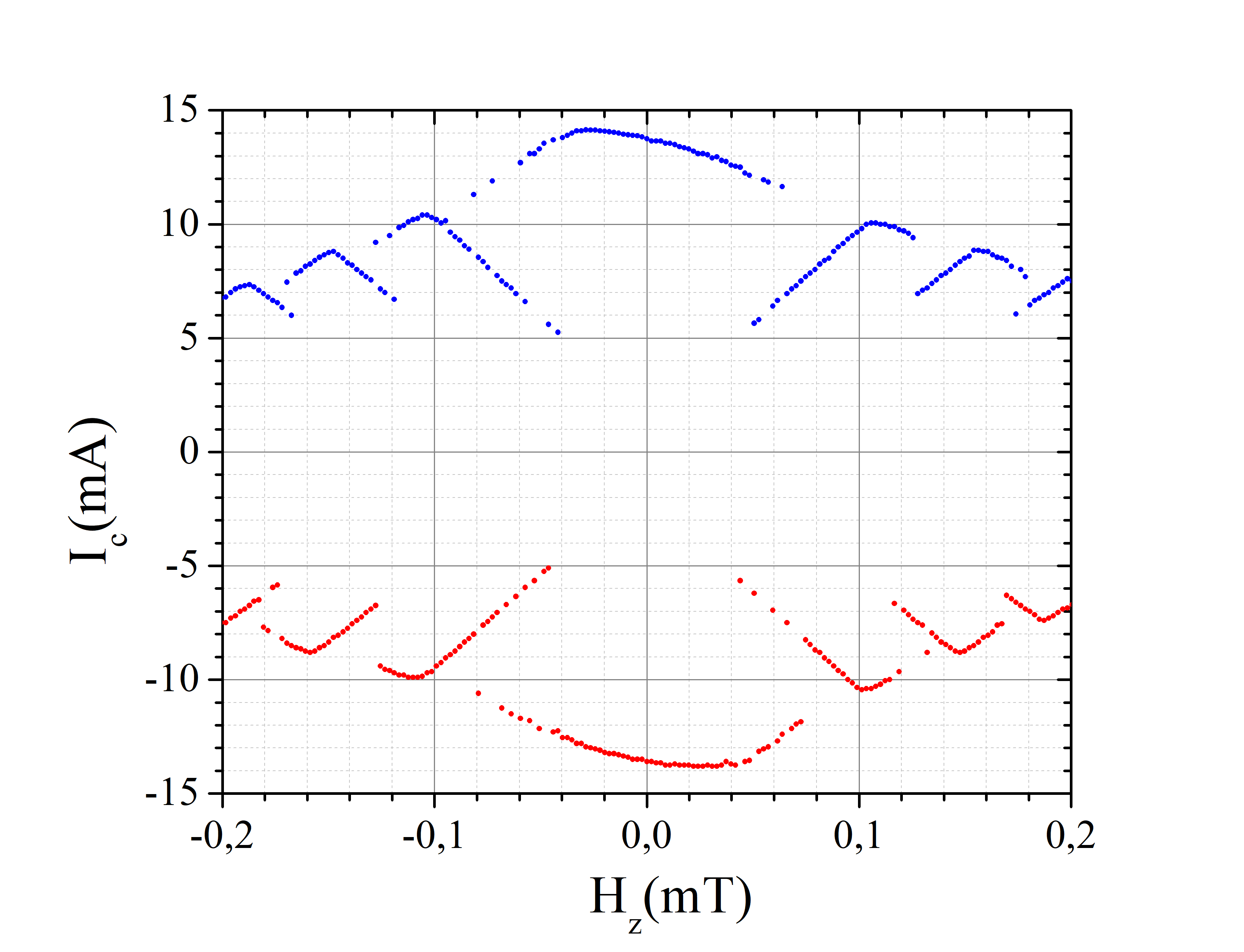}}
\subfigure[ ]{\includegraphics[width=7cm]{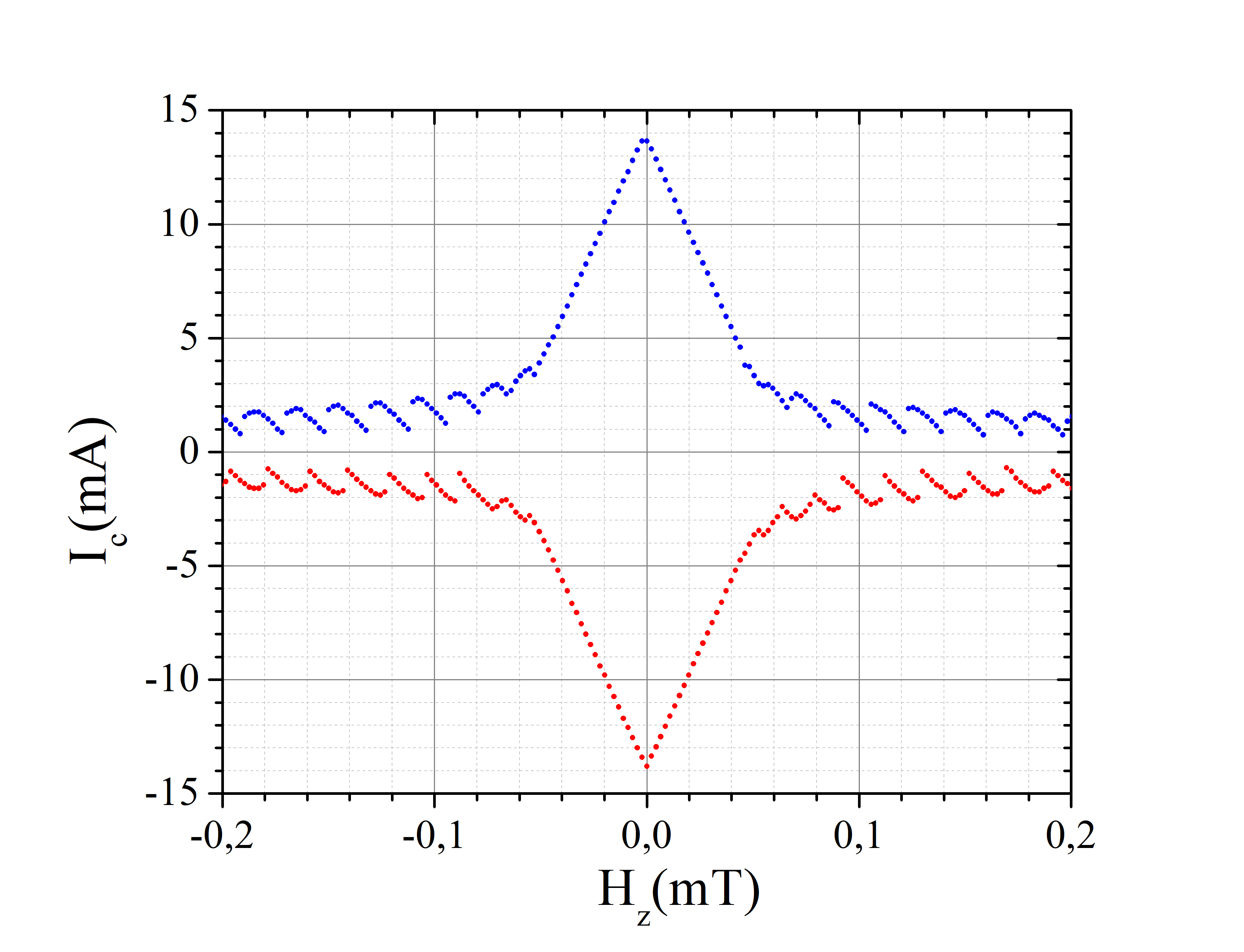}}
\subfigure[ ]{\includegraphics[width=7cm]{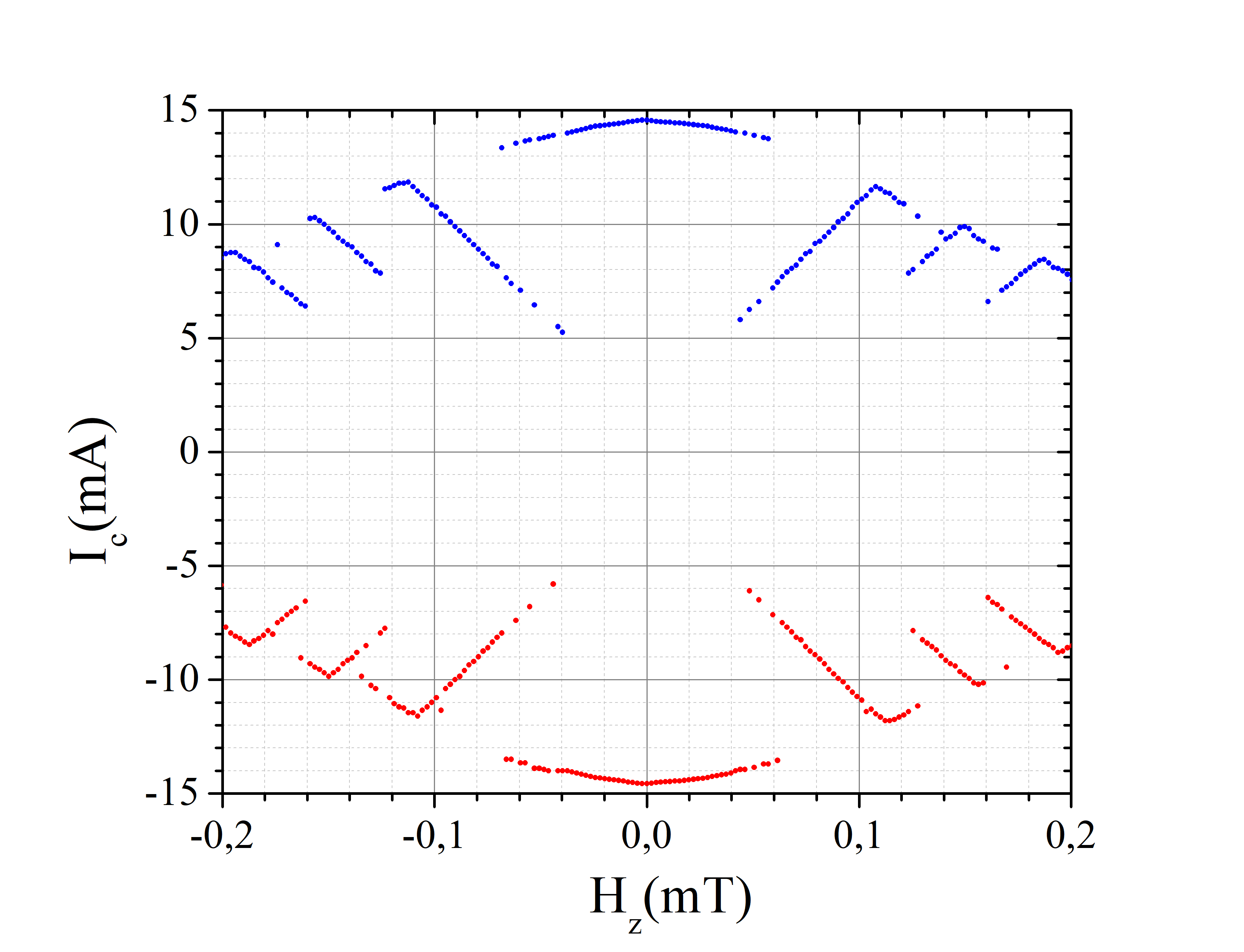}}
\caption{(Color online) Experimental threshold curves of our CAJTJs in a transverse magnetic field: (a) \jun A, (b) \jun B, (c) \jun C, (d) \jun D. The extrapolated dotted lines help to locate the critical fields.}
\label{perp}
\end{figure}

An alternative way to modulate the critical current of a planar \Jos tunnel \jun is to apply a magnetic field, $H_z$, perpendicular to the \jun plane\cite{rc,hf,miller,JAP08}, which induces shielding currents in its electrodes. In turn, the shielding currents generate a local magnetic field whose normal component thread the \Jos barrier. The modulation amplitude drastically depends on the geometry of the electrodes and on how close to the barrier the shielding currents circulate. It has been proven that this mechanism is particularly efficient in annular junction \cite{PRB09,SUST15}. Figures~\ref{perp}(a)-(d) display the $I_c$ vs. $H_z$ dependencies of the same \juns reported in Figures~\ref{par}(a)-(d). They can be interpreted according to the simple rule  that for Lyngby-type annuli a transverse field is equivalent to an-in plane field applied in the direction of the current flow \cite{SUST15}. In fact, while the transverse MDPs of junctions C and D are practically indistinguishable from their in-plane counterpart (apart from a field factor scale), those for junctions A and B are inverted. It is also seen that for our samples the transverse magnetic field is at least one order of magnitude more efficient than an in-plane field to modulate the critical currents. Later on, we will show that the fluxon state can be manipulated not only by an in-plane field but also by a transverse one.

\section{ Vortex state manipulation}

A goal of quantum information technology is to control the quantum state of a system, including its preparation, manipulation, and measurement. In this section we will investigate the state preparation and determination processes when one vortex has been trapped in a CAJTJ.

\subsection{ Vortex trapping}

\begin{figure}[tb]
\centering
\subfigure[ ]{\includegraphics[width=7cm]{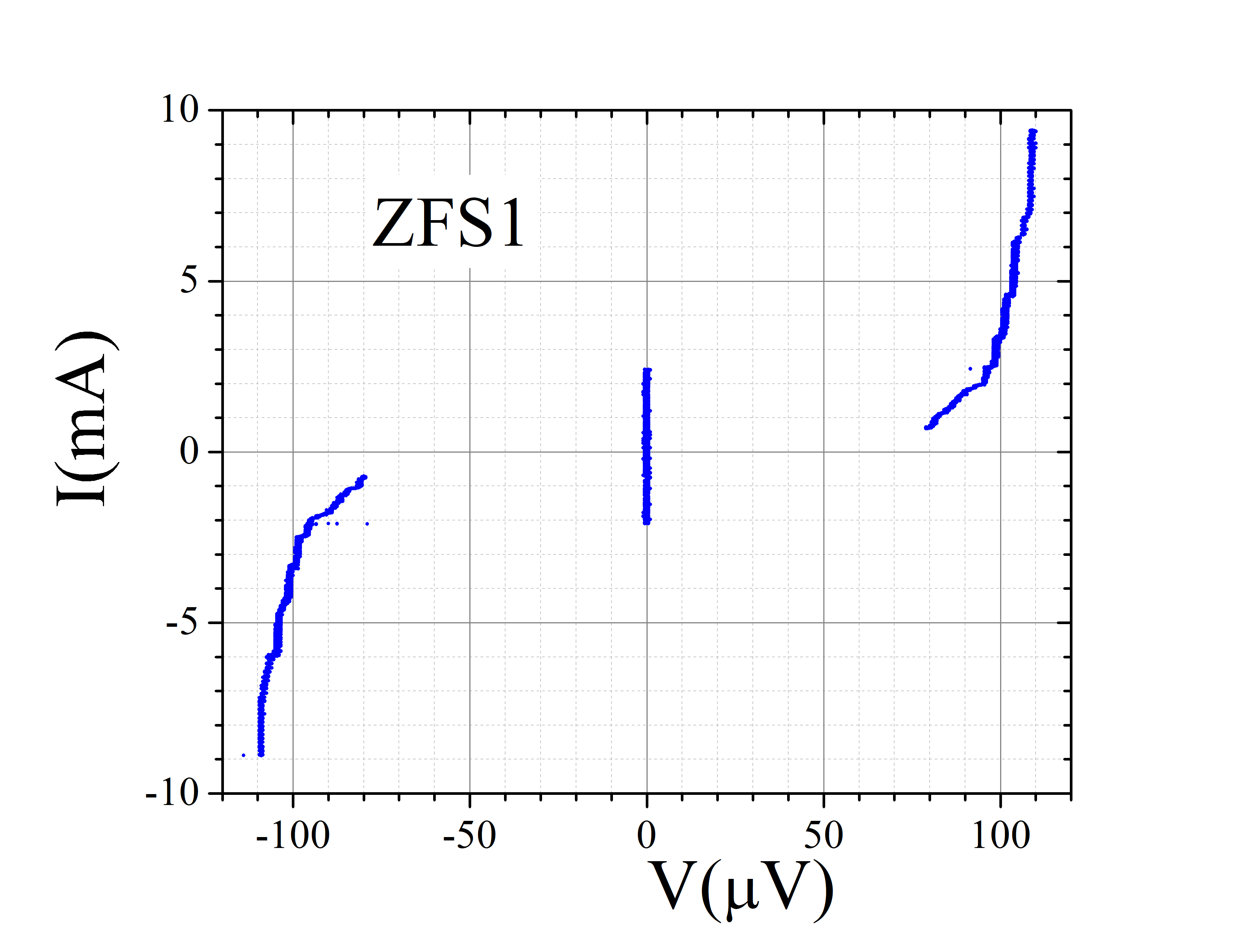}}
\subfigure[ ]{\includegraphics[width=7cm]{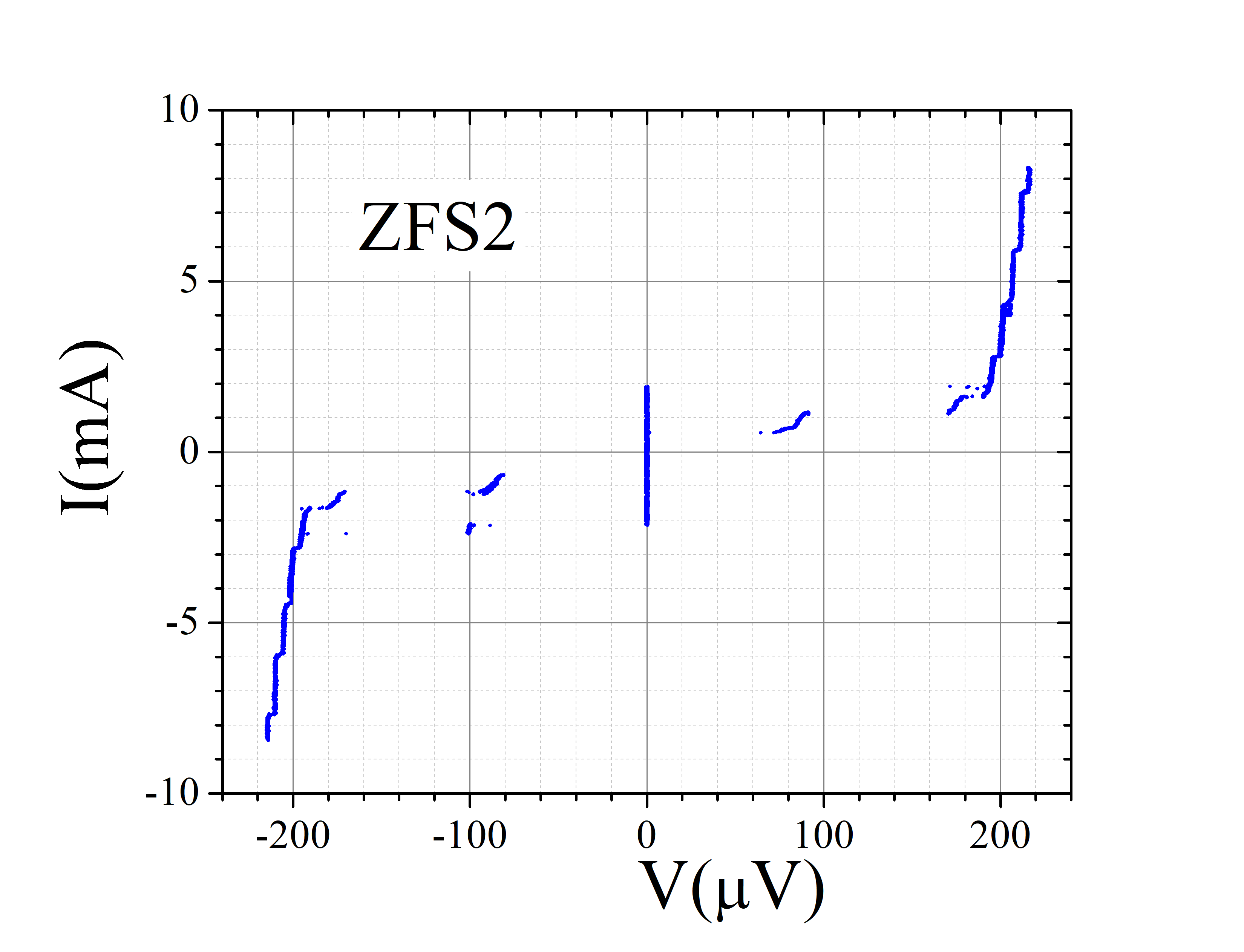}}
\caption{(Color online) Recorded current-voltage characteristics at $t=4.2\,K $ of a CAJTJ in the absence of an externally applied magnetic field with: (a) one fluxon trapped (ZFS1) and (b) two fluxons trapped (ZFS2).}
\label{ZFS}
\end{figure}

The spontaneous trapping of a magnetic flux in the superconducting loop formed by either the bottom or top electrode of a CAJTJ was achieved by repeatedly cooling the sample below the critical temperature of niobium, $T_c\approx 9.2 K$, with no bias current passing through the junction and no applied field. The chip was heated above the critical temperature by a voltage pulse applied to the integrated meander line heater. After the pulse the heat dissipates from the chip both through the thermal contact with the $Cu$ block and by the He exchange gas inside the can. At the end of each quenching cycle the possible spontaneously generated fluxons are static. An external current supplied to the CAJTJ sets the fluxons (if any) in motion around the annulus and quantized voltages develop across the junction itself. After a successful trapping attempt the number of trapped fluxons was determined from the voltage of the zero field step on the current-voltage characteristics. The trapping probability was found to be of about $10\%$  and not rarely two fluxons were trapped. Figures~\ref{ZFS}(a) and (b) show the profiles of, respectively, the first and second zero field step obtained by sweeping the bias current with a triangular waveform. The depinning of the fluxon(s) was observed as a switching from the zero voltage state at the current $\gamma_d$ that was smaller by a factor of about 5 than the critical current for the same junction, measured without trapped fluxons. This fact indicates the presence of a deep potential well (or multiple degenerate wells). For our samples the current branch associated with one fluxon had an asymptotic voltage $V_1 \approx 110\,\mu V$ which results in an average speed, $L V_1/ \Phi_0\approx 1.1 \times 10^7\, m/s$, considerably smaller than the Swihart velocity, $1.5 \times 10^7\, m/s$, typical of all-$Nb$ JTLs \cite{PRB03} evidencing, once again, that the fluxon travels in the periodic potential\cite{PRB98}. Well pronounced fine structures (generated by the resonant emission of plasma waves by the fluxon) appear in the ZFS profiles which progressively disappear as the temperature is increased. It means that at $4.2\,K$ the actual losses in the experiment are weaker than that taken for numerical simulations. Indeed, lower losses should be used in the simulations to enhance the fine structures. However, great care must be taken to simulate low damping nonlinear systems, since, besides the longer transients, the results are very sensitive to the numerical algorithm adopted to integrate the partial differential equation.

\subsection{ Vortex state preparation and determination}

\begin{figure}[t]
\centering
\subfigure[ ]{\includegraphics[width=7cm]{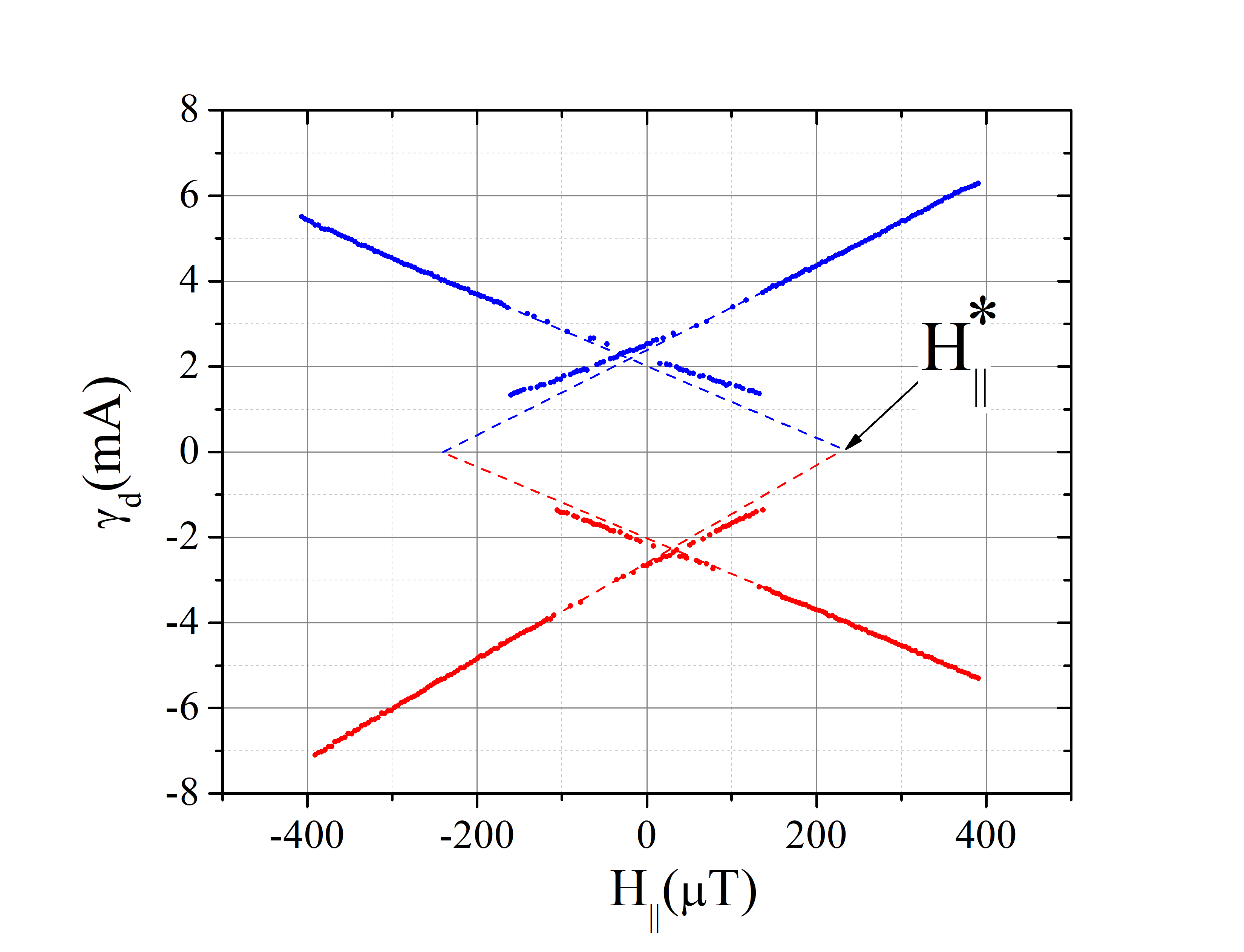}}
\subfigure[ ]{\includegraphics[width=7cm]{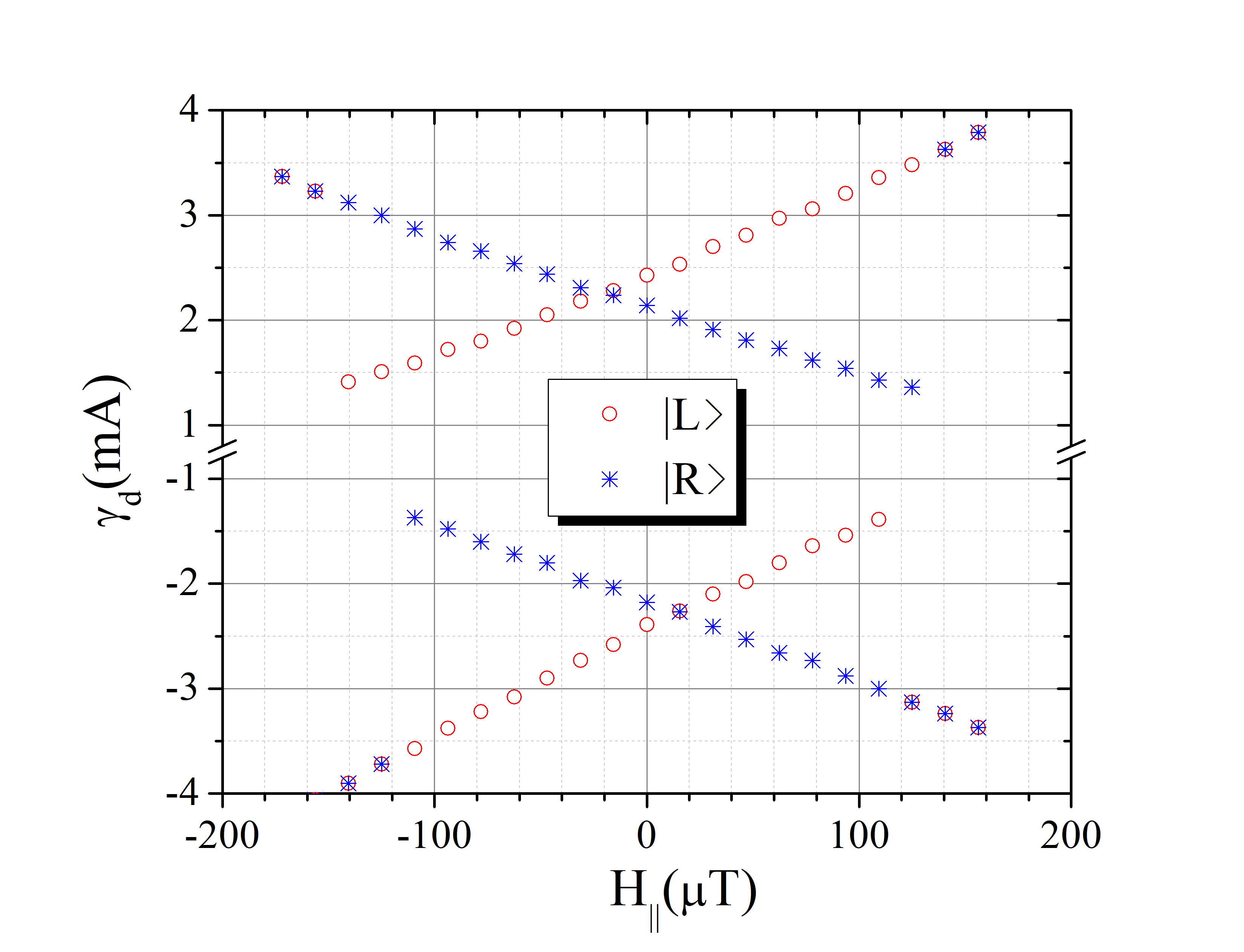}}
\caption{(Color online) Vortex-depinning current $\gamma_d$ as a function of applied in-plane magnetic field $H_{\parallel}$ . Measurements were recorded at a temperature of $T=4.2\,K$ using junction A.}
\label{depinn}
\end{figure}

\noindent Sometimes, upon applying a small magnetic field the depinning current becomes double-valued. This is a clear indication that bistable states exist and, when the sweeping current crosses zero, the decelerating fluxon can be trapped in two different potential wells; this is a statistical process that depends on the losses experienced by the fluxon and the relative depths of the potential wells. A neat example of this situation is given in Figure~\ref{depinn}(a) where the solid dots shows the positive and negative depinning currents in a parallel field measured by sweeping the bias current across junction A; double-valued depinning currents are clearly observed in a small magnetic range near the zero whose values can differ by as much as few milliamperes. The extrapolations of the almost linear branches help to locate the perpendicular threshold field $H_{||}^{*}\approx 200\,\mu T$, i.e., the smallest magnetic needed to prepare the fluxon state. Indeed these extrapolations can be done even in the more general cases when the double values are rare or even not observed by sweeping the IVC.

\begin{figure}[hb]
\centering
\subfigure[ ]{\includegraphics[width=7cm]{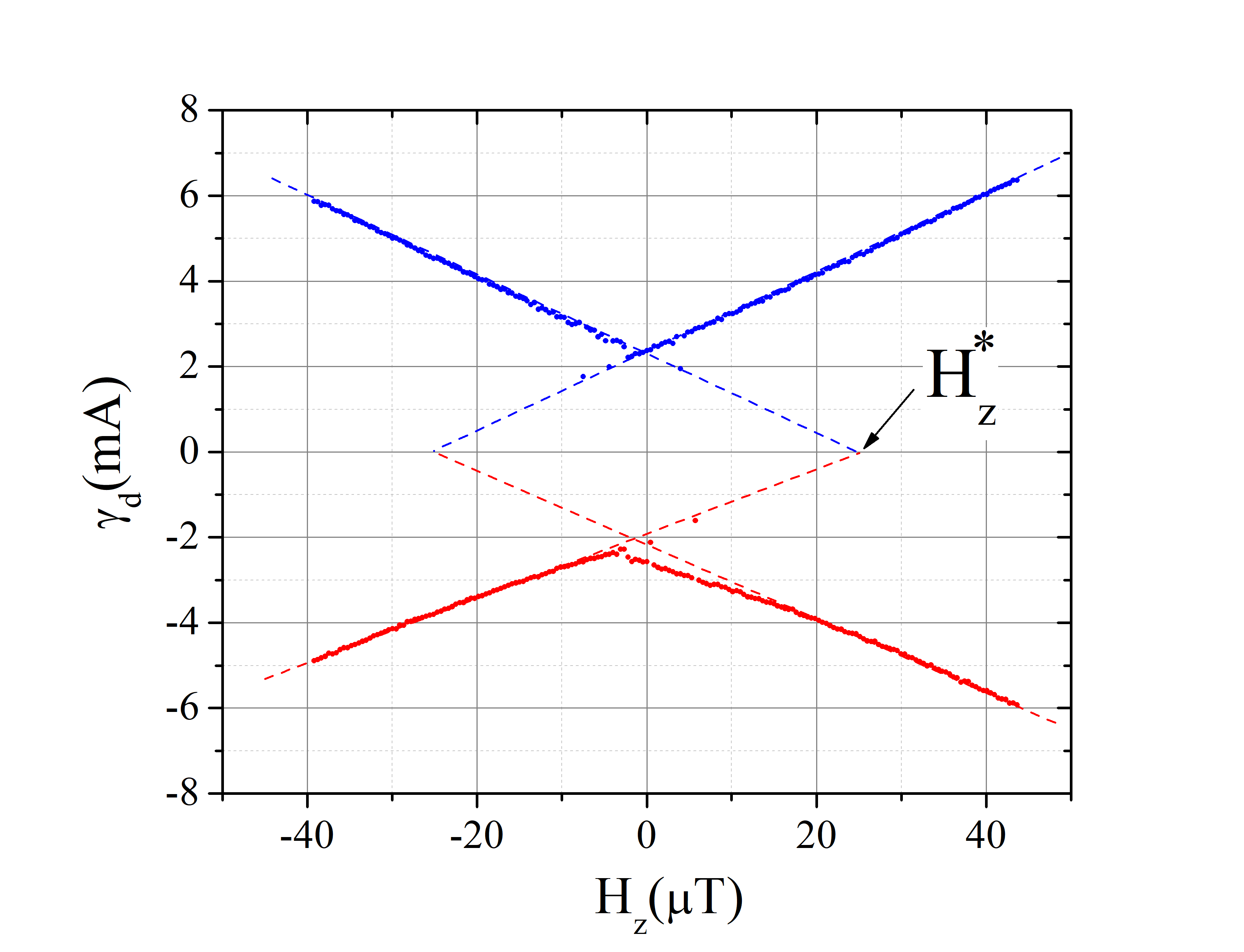}}
\subfigure[ ]{\includegraphics[width=7cm]{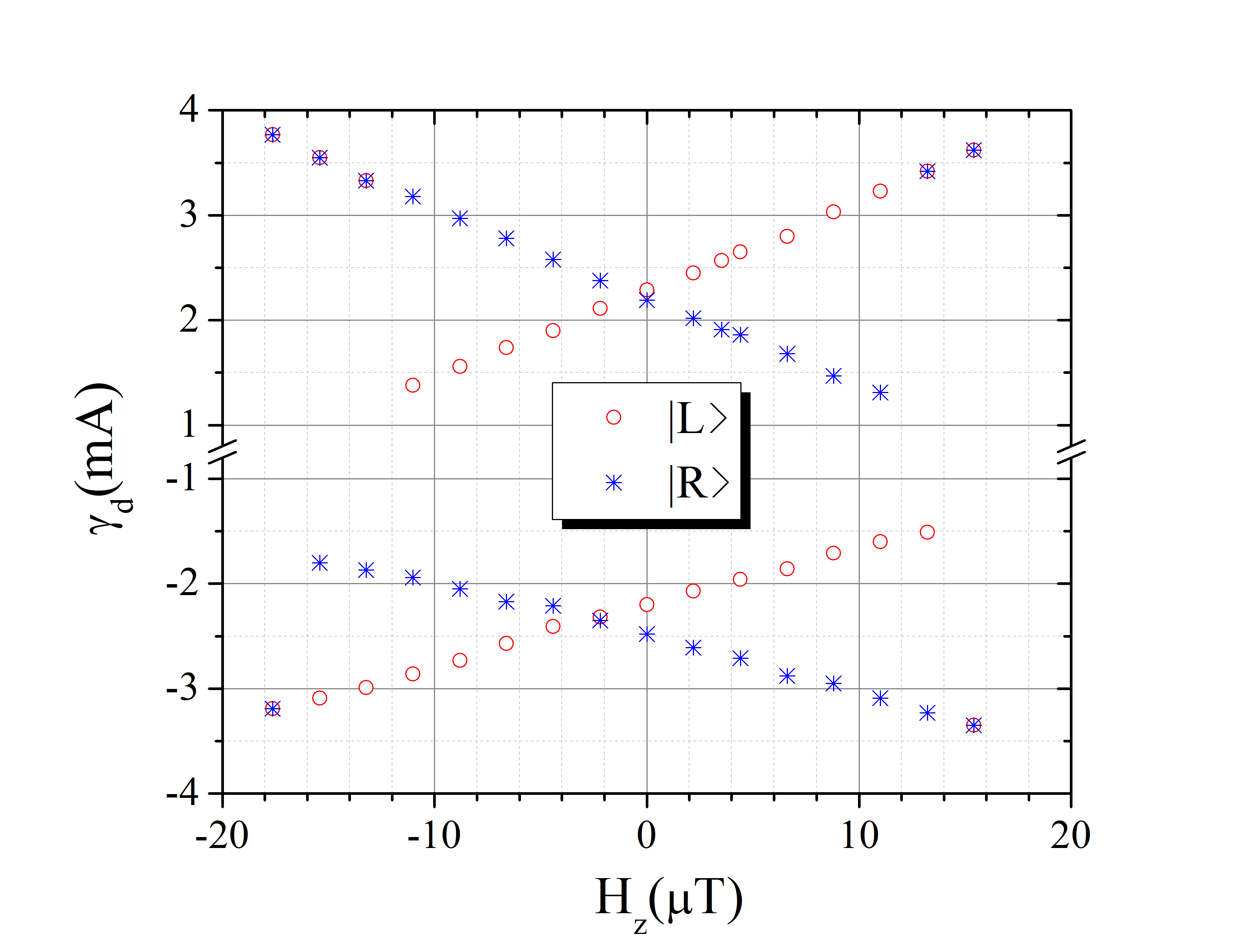}}
\caption{(Color online) Vortex-depinning current $\gamma_d$ as a function of applied transverse magnetic field $H_z$. Measurements were recorded at a temperature of $T=4.2\,K$ using junction A.}
\label{depinnZ}
\end{figure}
\noindent It turned out that the vortex state could be reliably prepared in either of the two possible states by simply applying to an unbiased CAJTJ a parallel magnetic field whose absolute strength exceeds $H_{||}^{*}$. The specific state depends on the polarity of the applied field, in full agreement with the numerical predictions of Section IIIB. Once prepared, the state of the static vortex can be read out by reducing the magnetic field in a small range near zero and increasing or decreasing the bias current until we observe a switch to a finite voltage. The result of this single-sweep measure of the depinning current unequivocally depends on the field polarity selected in the state preparation. The measured single-sweep positive and negative depinning currents are plotted as a function of the parallel magnetic field in Figure~\ref{depinn}(b). We conventionally named by $|L\rangle$ ($|R\rangle$) the state prepared by a field larger than $H_{||}^{*}$ (smaller than $-H_{||}^{*}$). Apart from a small tilting due to the self-fields, the data are quite symmetric and, as expected, $\gamma_{d-}(H_{\parallel}) \simeq -\gamma_{d+}(H_{\parallel})$. The only discrepancy with the predicted behavior is that the depinning currents merge well before the lowest one approaches zero, i.e., halving the effective field range that allows the unambiguous identification of the potential well where the resting vortex is located. Anyhow, small depinning currents correspond to shallow potential wells which can be smeared out by the thermal fluctuations that were not taken into account in the numerical simulations.

\noindent As predicted, the reliable preparation of the vortex state has been achieved also by means of a sufficiently large perpendicular field and, unexpectedly, the state determination has been found to be possible in a near-zero perpendicular field, meaning that when the fluxon is depinned from its original well it does not get trapped in the opposite well. This indicates that even in the presence of a perpendicular field the intra-well barrier of the current-tilted potential is comparable to the inter-well barrier and that the friction experienced by the fluxon is lower than that used in the numerical simulations. In different words, data qualitatively similar to those shown in Figures~\ref{depinn}(a) and (b) for a parallel field have been obtained in the presence of a perpendicular field. Moreover, the vortex state was found to be fully controllable even by means of a transverse magnetic field. The procedure is the same as for a parallel or perpendicular field and we will report the data for the same junction considered before in Figures~\ref{depinn}(a) and (b). First, as shown in Figure~\ref{depinnZ}(a), the transverse threshold field, $H_{z}^{*}$, is evaluated by recording the magnetic field dependence of the depinning current while continuously sweeping on the junction IVC; we see that the double-solutions are rare in this specific case. Later on, as shown in Figure~\ref{depinnZ}(b), the single-sweep depinning currents are measured after a preparation stage in which a large, either positive or negative, transverse field is applied to the unbiased junction. The transverse field has the advantage to be about ten times more efficient.

\section{Conclusions}
\vskip -8pt

We have studied a vortex qubit based on an annular Josephson tunnel junction delimited by two closely spaced confocal ellipses that is characterized by a periodically modulated width. This spatial dependence, in turn, generates a periodic potential that alternately attracts and repels the fluxons (or antifluxons). The potential energy minima occur at two diametrically opposite locations where the annulus is narrowest and the intra-well potential height is uniquely determined by the CAJTJ aspect ratio. This configuration is faithfully modeled by a modified and perturbed one-dimensional sine-Gordon equation that admits (numerically computed) solitonic solutions.

\noindent The proposed vortex qubit design has been tested experimentally in the classical regime and bistable vortex states were observed on high-quality $Nb/Al$-$AlOx/Nb$ CAJTJs having an aspect ratio $\rho=0.5$. Preparation of the vortex in a given potential well was achieved by means of an external magnetic field of proper polarity applied either in the barrier plane or in the transverse direction. The final state of the vortex can be read out by performing an escape measurement from one of the potential wells in the presence of a small magnetic field. In our experiments carried out at $T=4.2\,K$ the fluxon escapes from a well in the tilted potential by a thermally activated process. At lower temperatures thermal activation as well as dissipation processes are exponentially suppressed, and the magnetic field range that allows the determination of the fluxon state is expected to widen. The transition from the thermal to the quantum regime, already observed in some Josephson junction systems, was typically found around $200\,mK$. Below this crossover temperature the quantum nature of the fluxon manifests as quantized energy levels within each potential well and the fluxon escape occurs by macroscopic quantum tunneling. Under sufficient decoupling from the environment, as with other superconducting qubits, the superposition of the macroscopically distinct states $|L\rangle$ and $|R\rangle$, not yet observed for Josephson vortex qubits, could be identified by means of the analysis of the switching current probability distribution and employed to implement a reliable Josephson vortex qubit.

\section*{Acknowledgments}
\noindent RM and JM acknowledge the support from the Danish Council for Strategic Research under the program EXMAD. VPK acknowledges support from the grant no.8168.2016.2 within the State Program for Support of Leading Scientific Schools .

\newpage

\end{document}